\documentclass[fleqn,usenatbib]{mnras}

\usepackage[T1]{fontenc}

\DeclareRobustCommand{\VAN}[3]{#2}
\let\VANthebibliography\thebibliography
\def\thebibliography{\DeclareRobustCommand{\VAN}[3]{##3}\VANthebibliography}

\usepackage{graphicx}	
\usepackage{amsmath}	
\usepackage{amssymb}	
\usepackage{newtxtext,newtxmath}

\usepackage{soul,xcolor}
\usepackage[normalem]{ulem}

\newcommand\rst{\bgroup\markoverwith{\textcolor{violet}{\rule[0.5ex]{2pt}{0.4pt}}}\ULon}

\title[The Age-Metallicity Relation]{Constraining the solar neighbourhood age-metallicity relation from white dwarf-main sequence binaries}

\author[A. Rebassa-Mansergas et al.]{A. Rebassa-Mansergas$^{1,2}$\thanks{E-mail: alberto.rebassa@upc.edu}, J. Maldonado$^{3}$, R. Raddi$^{1}$, A. T. Knowles$^{1}$, S. Torres$^{1,2}$, M. Hoskin$^{4}$, \newauthor  T. Cunningham$^{4}$, M. Hollands$^{4}$, J. Ren$^{5}$, B. T. G\"ansicke$^{4}$, P.-E. Tremblay$^{4}$, N. Castro-Rodr\'iguez$^{6,7,8}$\newauthor M. Camisassa$^{9,1}$, D. Koester$^{10}$
\\
$^{1}$ Departament de F\'isica, Universitat Polit\`ecnica de Catalunya, c/Esteve Terrades 5, 08860 Castelldefels, Spain\\
$^{2}$ Institute for Space Studies of Catalonia, c/Gran Capit\`a 2--4, Edif. Nexus 104, 08034 Barcelona, Spain\\
$^{3}$ INAF - Osservatorio Astronomico di Palermo, Piazza del Parlamento 1, 90134 Palermo, Italy\\
$^{4}$ Department of Physics, University of Warwick, Coventry, CV4 7AL, UK\\
$^{5}$ CAS Key Laboratory of Space Astronomy and Technology, National Astronomical Observatories, \\Chinese Academy of Sciences, Beijing 100101, China\\
$^{6}$ Instituto de Astrof\'isica de Canarias, E-38200, La Laguna, Tenerife, Spain\\
$^{7}$ Universidad de La Laguna. Depart. de Astrof\'isica, E-38206, La Laguna, Tenerife, Spain\\
$^{8}$ GRANTECAN, Cuesta de San Jos\'e s/n, E-38712 , Bre\~na Baja, La Palma, Spain\\
$^{9}$ Department of Applied Mathematics, University of Colorado, Boulder, CO 80309-0526, USA\\
$^{10}$ Institut f\"ur Theoretische Physik und Astrophysik, Universit\"t Kiel, 24098, Kiel, Germany
}

\date{Accepted XXX. Received YYY; in original form ZZZ}

\pubyear{2021}

\begin{document}
\label{firstpage}
\pagerange{\pageref{firstpage}--\pageref{lastpage}}
\maketitle

\begin{abstract}
The age-metallicity  relation is  a fundamental tool  for constraining
the chemical evolution  of the Galactic disc. In this  work we analyse
the observational properties of this  relation using binary stars that
have not  interacted consisting of a  white dwarf --from which  we can
derive the total  age of the system-- and a  main sequence star --from
which  we  can  derive  the   metallicity  as  traced  by  the  [Fe/H]
abundances.   Our   sample  consists  of  46   widely  separated,  but
unresolved spectroscopic binaries identified  within the Sloan Digital
Sky  Survey, and  189 white  dwarf  plus main  sequence common  proper
motion pairs identified within the second data release of \emph{Gaia}.
This  is  currently the  largest  white  dwarf  sample for  which  the
metallicity of their progenitors have been determined.  We find a flat
age-metallicity relation displaying a  scatter of [Fe/H] abundances of
approximately $\pm$0.5 dex  around the solar metallicity  at all ages.
This independently  confirms the lack  of correlation between  age and
metallicity  in the  solar  neighbourhood that  is  found in  previous
studies  focused on  analysing  single main  sequence  stars and  open
clusters.
\end{abstract}




\section{Introduction}

The formation and assembly history of our Galaxy can be traced through
the  chemical   composition,  age,   position  and  velocity   of  its
constituent stars. In particular,  the age-metallicity relation (AMR),
that  is the  observed connection  between  the age  and the  chemical
abundances of  stars, can provide  vital constraints on  the formation
and  evolution  of the  Galactic  disc.  The  AMR has  therefore  been
widely-studied  during  the last  decades  (e.g.  see the  reviews  of
\citealt{Nomoto13}  and  \citealt{Feltzing13}   and  the  more  recent
studies by \citealt{Bergemann14} and \citealt{Wojno2018}). Some of the
very early works on this topic found  evidence for an AMR in which the
oldest  stars have  the  lowest  metallicity (e.g  \citealt{Twarog80},
\citealt{RochaPinto2000},  \citealt{Soubiran08}).  This  relation  was
found to behave  as expected in a formation and  evolution scenario in
which  stars  form  from  the  metal-enriched  gas  left  by  previous
generations of stars. However, there have been evidence against an AMR
existing  at   all,  with  large  metallicity   dispersions  found  in
comprehensive  or   more  recent   works  \citep[e.g.][]{Edvardsson93,
  Haywood13, Bergemann14, Wu2021}. The lack of correlation between age
and  metallicity in  the Milky  Way suggests  a much  more complicated
formation scenario \citep{Feuillet2019}.

A  major limitation  of most  previous studies  is the  measurement of
stellar ages,  which is  a difficult task.  Historically, the  ages of
stars   were   estimated   through   chromospheric   activity   levels
(e.g.   \citealt{Barry88};    \citealt{RochaPinto2000})   or   through
isochrone     fitting     with      an     appropriate     metallicity
(e.g.  \citealt{Jorgensen05}; \citealt{Howes19}).  More recently,  the
combination  of spectroscopy  and  asteroseismology has  been used  to
estimate    stellar    masses    and   to    derive    stellar    ages
(e.g. \citealt{Casagrande2016, Pinsonneault18}).

An alternative method  to estimate the age component of  the AMR is to
use white dwarfs (WDs) as cosmic clocks. WDs are the typical end stage
of the  vast majority of main  sequence (MS) stars (see  the review of
\citealt{Althaus2010}),   and  because   their  evolution   follows  a
relatively simple  and well  understood cooling  process, they  can be
used   as    reliable   observational   measures   of    stellar   age
\citep[e.g.][]{Fouesneau2019, Qiu2020, Lam2020}. In order to determine
the total age of  a WD, defined as the sum of its  cooling age and its
MS  progenitor   lifetime,  two  processes  are   required.  First,  a
prescription for evolutionary cooling  sequences provides a measure of
the  WD   cooling  age  from  observed   determinations  of  effective
temperature ($T_\textrm{eff}$) and surface  gravity (log $g$). To that
end  we  adopt  the  widely-used  sequences  of  the  La  Plata  group
(e.g.      \citealt{Renedo2010,       Althaus2015,      Camisassa2016,
  Camisassa2019}), which encompass the full range of WD masses and the
most   updated   prescriptions   in    the   treatment   of   physical
processes. Second, a  relationship between the mass of the  WD and the
mass  of  its   progenitor  is  required  to  obtain   a  MS  lifetime
estimate.       This        initial-to-final       mass       relation
(e.g. \citealt{Catalan08,  Cummings2018, Barrientos2021})  enables the
estimation  of MS  star masses  from measures  of the  current day  WD
masses, which in turn, can be used  to determine the time spent on the
MS   from  evolutionary   sequences,  provided   the  metallicity   is
known. Once we know the MS  lifetimes and the cooling times, the total
ages are determined as the sum of these two ages.

Because a  significant fraction of  known MS  stars in our  Galaxy are
expected  to  be  in  multiple systems,  particularly  binaries  (e.g.
\citealt{Yuan15}),  and because  the majority  of MS  stars end  their
lives as  WDs, it follows that  there exists a large  number of binary
systems consisting of  a primary (more massive) star  that has evolved
to a WD,  and a secondary MS  star. The majority of  these WDMS binary
systems are expected to have  orbital separations that are wide enough
to avoid mass transfer episodes, thus allowing one to treat them as if
they  were isolated  (\citealt{Willems04}).   These  wide WDMS  binary
systems can therefore  be considered as useful probes in  the study of
the Galactic AMR. An  accurate age of the WD can  be estimated via the
process described  above and the  MS star metallicity can  be measured
via   detailed  spectroscopic   matching   to   stellar  models   (e.g
\citealt{Bensby2014}; \citealt{Holtzman2015}). Assuming  the two stars
within the binary  are coeval, a measurement of the  WD age also gives
an age estimate for the MS star,  and thus the AMR can be probed using
these  systems.   This work  follows  on  from  the initial  study  of
\citet{Rebassa16AMR},  in which  an AMR  was determined  from a  pilot
sample  of 23  wide WDMS  binary systems  identified within  the Sloan
Digital Sky  Survey (SDSS; \citealt{York2000,  Eisenstein2011}). Here,
we continue  this observational campaign and  considerably enlarge the
sample  using WDMS  binaries in  resolved common  proper motion  pairs
(CPMPs) identified thanks to the \emph{Gaia} mission \citep{Gaia2016}.

The structure  for this  paper is as  follows. Section\,\ref{s-sample}
presents    the     WDMS    binary     samples    used     for    this
work. Section\,\ref{s-obs} describes the  observations of both WDs and
MS  stars. Section\,\ref{s-methods}  presents the  results, which  are
discussed       in        Section\,\ref{s-discussion}.        Finally,
Section\,\ref{s-concl} presents our conclusions.

\section{The WDMS Binary Samples}
\label{s-sample}

In this work we use two different WDMS binary samples to constrain the
age-metallicity relation.

The first  is the catalogue of  more than 3000 WDMS  binaries from the
SDSS \citep{Rebassa2016b}.  In particular,  for this project we select
binaries containing  a hydrogen-rich white dwarf  primary that display
no significant radial velocity variations  over a baseline of at least
two  days  and  have  SDSS   $g$  magnitudes  $<$19\,mag.   The  first
requirement ensures  the binaries  are widely separated  and therefore
very likely  did not evolve through  any phase of mass  transfer (note
that the orbital period distribution of close SDSS WDMS binaries peaks
at $\simeq$8  hours and that  there are  very few close  binaries with
periods  longer  than 1  day,  see  \citealt{Nebot2011}).  The  second
condition  selects systems  that are  bright enough  to allow  for the
derivation  of  reliable  WD  ages  and  MS  star  metallicities  from
spectroscopic observations  at ground-based  telescopes of  medium and
large apertures.   In \citet{Rebassa16AMR} we presented  a pilot study
of the age-metallicity relation from 23 SDSS WDMS binaries and here we
extend the analysis to 29 additional systems.

We build our second WDMS binary sample by mining the data release 2 of
\emph{Gaia}  \citep{Gaia2016,  Gaia2018}.   Our approach  here  is  to
identify CPMPs  that contain a  WD and a MS  star. This has  two clear
advantages over the SDSS WDMS binary sample introduced in the previous
paragraph. Firstly, the WD and MS  stars are separated enough to allow
for  independent observations  of  the two  stellar components.   This
avoids any  possible selection  effects against the  identification of
binaries containing cool WDs, which  are generally outshined by the MS
companions in the  SDSS sample \citep{Rebassa2016b}. These  WDs are of
great interest since they have the  longest cooling ages and can hence
be used to  test the age-metallicity relation  at the intermediate/old
ages. Secondly, the MS companions are not only M dwarfs, which are the
dominant spectral types in the SDSS  WDMS binary sample, but also F, G
and K-type  stars. M-dwarf metallicity  calibrators that are  based on
the  spectral  analysis  have  been often  criticised.   For  example,
\citet{Lindgren2017}  argue that,  for individual  M stars,  different
metallicity    calibrations   yield    [Fe/H]\footnote{[Fe/H]=   $\log
  \left(\frac{\mathrm{nFe}}{\mathrm{nH}}\right)_{*}       -       \log
  \left(\frac{\mathrm{nFe}}{\mathrm{nH}}\right)_{\odot}$,        where
  $\left(\frac{\mathrm{nFe}}{\mathrm{nH}}\right)$   is    the   number
  abundance ratio of Fe relative to H  for a given star (*) or for the
  Sun ($\odot$).}  abundances that differ by  as much as 0.6 dex.  The
\emph{Gaia} WDMS binary sample containing F, G and K companions allows
us  to  avoid this  possible  issue,  since  it  has been  shown  that
metallicity  calibrations provide  similar values  of [Fe/H]  for such
stars \citep[e.g.][]{Teixeira2016}.

\begin{figure}
    \centering
    \includegraphics[width=\linewidth]{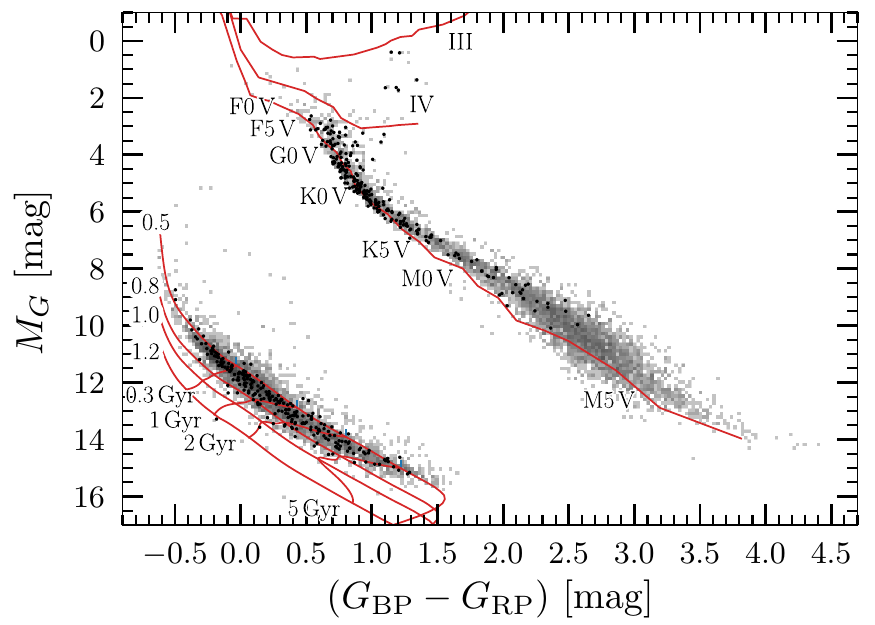}
    \caption{Hertzsprung-Russell diagram  of {\em  Gaia}-selected WDMS
      common proper-motion pairs.  The  dwarf (V), sub-giant (IV), and
      giant  (III)  tracks  are obtained  from  the  \citet{pickles98}
      spectral library. The cooling sequences for hydrogen-rich WDs of
      0.5,   0.8,  1.0,   and   1.2\,M$_{\odot}$   are  described   by
      \citet{Camisassa2016} and  \citet{Camisassa2019}. The isochrones
      for total WD ages, that is  the pre-WD evolutionary time and the
      WD cooling  age, are also shown  for 0.3, 1, 2,  and 5\,Gyr. The
      black dots  represent the companion stars  with available [Fe/H]
      abundances  from  our  spectroscopic   follow-up  and  their  WD
      primaries.}
    \label{fig:hr}
\end{figure}

We searched  for CPMP companions  to the  WDs that were  identified by
\citet{gentile-fusillo2019} in  {\em Gaia} DR2. From  their catalogue,
we selected 103,002 candidates with a probability of being a WD larger
than   50~percent  and   a   parallax  uncertainty   of  better   than
10~percent. Thus,  we mined the  {\em Gaia} DR2 archive  searching for
common  proper motion  MS  companions within  100,000~au  from the  WD
candidates.   Our  procedure   mimics  that  of  \citet{el-badry2018},
although  we used  an input  WD  list and  we adopted  a wider  search
radius.  By  assuming the  two stars  are on  a bound  Keplerian orbit
\citep[cf.][]{andrews2017,   el-badry2018,  jimenez-esteban2019},   we
impose an  upper limit of  8.5\,M$_{\sun}$ for  the total mass  of the
system, which implies a maximum  difference in projected velocities of
$\Delta       V_\perp$       (km/s)      $\leq       2.73       \times
(a/10^{3}~\mathrm{au})^{-1/2}$, where $a$  is the projected separation
in   astronomical  units.    We  applied   standard  astrometric   and
photometric cuts  \citep{gaia-hr2018, lindegren2018-b} to  ensure that
companions with good quality data are found.  Hence, we identify 4,415
WDMS   systems  with   distances  $\la$500\,pc,   with  a   log-normal
distribution  centered at  $\log$(1/$\varpi$)  = 2.17$\pm$0.25,  where
0.25 is the standard deviation  and $\varpi$ is the measured parallax.
A comparison with the catalogue of \citet{el-badry2018} shows that our
search  resulted  in  $\approx  1800$ additional  systems,  while  our
catalogue does not contain 194  systems found by these authors.  These
differences are mostly explained by the use of an input WD list, which
enables us to  expand our catalogue towards larger  distances from the
Sun.  A  comparison with  the most recent  catalogue of  common proper
motion pairs identified by  \protect\citet{el-badry2021} confirms a 92
percent overlap, with  the majority of pairs having  a low probability
of   chance    alignment   as    defined   by   the    authors.    The
Hertzsprung-Russell diagram  of our resolved CPMP  candidates is shown
in Fig.\,\ref{fig:hr}.

\section{Observations of the WDMS binary sample}
\label{s-obs}

We adopted different strategies to observe the WDMS binaries belonging
to the  two samples described  in the previous  section. A log  of the
observations is shown in Table\,\ref{t-log}.

SDSS WDMS binaries  are unresolved despite the fact  that they display
no  radial velocity  variations. This  is because  they are  generally
located    at    far    enough    distances    \citep[$\simeq$400--500
  pc;][]{Rebassa2010}.  Therefore, one  single  spectrum collects  the
dispersed light from both components.

The \emph{Gaia} WDMS binaries we selected  in this work are members of
spatially resolved CPMPs. As a  consequence, we aimed at obtaining one
optical spectrum for each component.

\subsection{SDSS WDMS binaries}

\subsubsection{Very Large Telescope}

We observed 17 SDSS WDMS binaries with the Very Large Telescope (VLT),
at   Cerro    Paranal   (Chile)    and   the    X-Shooter   instrument
\citep{Vernet2011}.    X-Shooter  provides   spectra  in   three  arms
simultaneously      covering      a      wavelength      range      of
$\simeq$3,000--25,000\AA\,   (UVB  arm;   3,000--5,600\AA,  VIS   arm;
5,500--10,200\AA, NIR  arm; 10,200--24,800\AA). The  observations were
performed in service  mode using the 0.9--1" slits,  which resulted in
spectra  with resolving  power  4,350/7,450/5,300  in the  UVB/VIS/NIR
arms,  respectively. We  reduced  and calibrated  the  data using  the
\textsc{esoreflex} X-Shooter pipeline, version 3.3.5.

\begin{table}
\caption{Log  of the  observations including  the telescope  name, the
  observing  mode (service,  sm, or  visitor, vm)  the month  dates or
  period of  the observations, the  instrument used and the  number of
  taken spectra.}  \setlength{\tabcolsep}{0.3ex}
\begin{center}
\begin{tabular}{ccccc}  
\hline \hline 
Telescope & Mode & Dates or period & Instrument & \#spec \\
\hline 
VLT          & sm &  P101& X-Shooter &  17 \\
GTC          & sm & 2016B, 2017A & OSIRIS    &  10 \\
GTC          & sm & 2017B, 2018A & EMIR      &  12 \\
WHT          & vm &  Oct. 2018, Feb. \& Apr. 2019   & ISIS & 122\\
INT          & vm & Jul. 2019 & IDS       &  92 \\
TNG          & sm & 2018B & HARPS-N  & 106\\
Mercator     & vm & Dec. 2018 \& Jan. 2019 & HERMES    &   377 \\
Xinglong2.16 & vm &  Dec. 2018, Feb. 2019 \& June 2020 & echelle   &  51 \\
\hline
\end{tabular}
\end{center}
\label{t-log}
\end{table}

\subsubsection{Gran Telescopio Canarias}
\label{s-gtc}

We obtained  service mode optical  and near-infrared spectra  with the
Gran Telescopio  Canarias (GTC),  for 10 and  12 additional  SDSS WDMS
binaries using the OSIRIS and  EMIR instruments, respectively. We used
the 2,000B and 2,500R gratings together with the 0.6 arcsec slit width
for the optical OSIRIS observations, which resulted in optical spectra
covering the 3,960-5,690\AA\, and 5,590-7680\AA\, wavelength ranges at
resolving  powers of  2,000  and 2,500.   The  EMIR observations  were
performed  using the  K  grism and  the 0.8  arcsec  slit width,  thus
providing spectra covering the 20,300--23,850\AA\, wavelength range at
a resolving power of 3,100. The  OSIRIS spectra were reduced using the
\textsc{pamela}  software   \citep{Marsh1989}  and   calibrated  using
\textsc{molly}\footnote{The \textsc{molly} package is developed by Tom
  Marsh            and            is           available            at
  http://deneb.astro.warwick.ac.uk/phsaap/software.}. The EMIR spectra
were reduced and calibrated using \textsc{redemir}, a new GTC pipeline
written in python.  In a first  step it eliminates the contribution of
the  sky background  using consecutive  A-B pairs.   Subsequently, the
spectra  are flat-fielded,  calibrated in  wavelength and  combined to
obtain the final spectrum in the K band.

\subsection{\emph{Gaia} WDMS CPMPs}

\subsubsection{William Herschel Telescope}

Low-resolution  spectra for  122 WDs  were obtained  at the  4.2-meter
William  Herschel  Telescope  (WHT)  at  El  Roque  de  los  Muchachos
observatory in La Palma in visitor  mode during 2018 and 2019. We used
the  ISIS (Intermediate-dispersion  Spectrograph  and Imaging  System)
instrument  and  the  1  arcsec  width  long-slit  together  with  two
different  gratings,  the 600B  and  600R.   This provided  blue-  and
red-arm simultaneous spectra at a resolving power of 600, respectively
covering  the $\simeq$3,600--5,100\AA\,  and $\simeq$5,600--7,200\AA\,
wavelength ranges. Arc spectra were  taken along the nights to account
the  flexure of  the instrument.   The  WHT spectra  were reduced  and
calibrated  using the  \textsc{pamela}  and \textsc{molly}  softwares,
respectively.

\begin{figure*}
\includegraphics[angle=-90,width=0.8\textwidth]{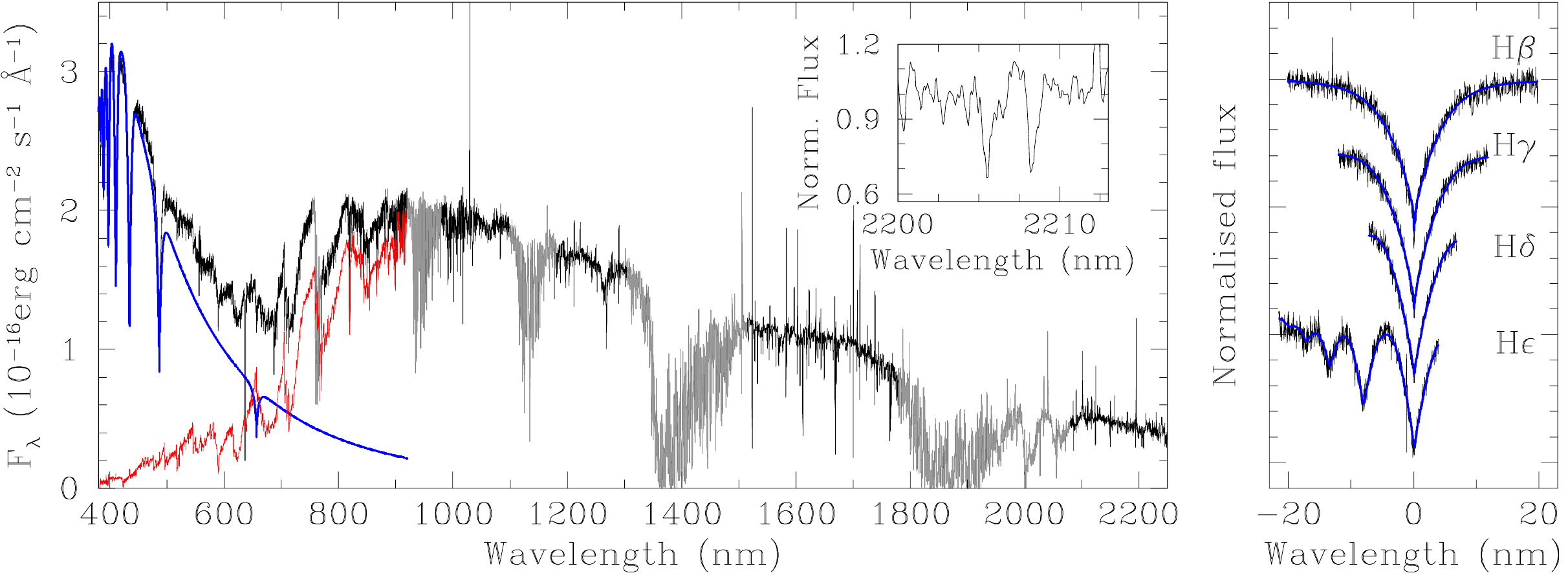}
\includegraphics[angle=-90,width=0.8\textwidth]{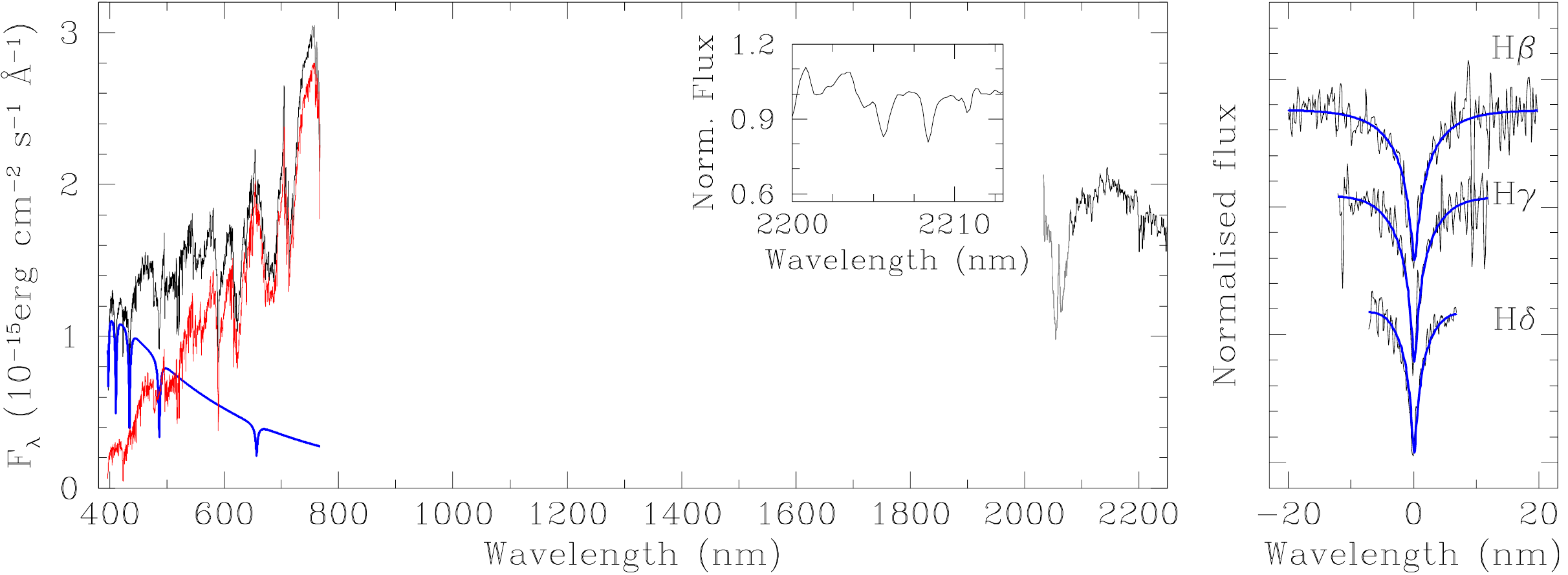}
\caption{Top-left  panel: spectrum  of SDSSJ\,0335+0038  (black) taken
  with X-Shooter. The  telluric absorption regions are  shown in grey.
  At  blue wavelengths  ($\lambda<$600  nm) the  flux contribution  is
  dominated by the  WD. Conversely, the red  and near-infrared regions
  are dominated by the flux pf the MS companion. In blue and red solid
  lines  we show  the  best-fitting  WD model  and  MS star  template,
  respectively. The top right corner shows the 2205/2209 nm Na~{\sc i}
  absorption doublet  (free of  telluric absorption) used  to estimate
  the [Fe/H] abundance of the MS star. Top-right panel: the normalized
  residual  and  normalised WD  Balmer  lines  (black) resulting  from
  subtracting the MS star together with  the WD model (blue) that best
  fits   the   spectrum.    Bottom    panels:   the   same   but   for
  SDSSJ\,1510+4048, which was observed with OSIRIS and EMIR mounted at
  the GTC.   Note that the EMIR  spectrum in the near-infrared  is not
  flux  calibrated and  has  been normalised  to  the average  optical
  flux.}
\label{f-spec}
\end{figure*}

\subsubsection{Isaac Newton Telescope}

Low-resolution spectra of 92 additional WDs were obtained at the Isaac
Newton Telescope  (INT) at  El Roque de  los Muchachos  observatory in
visitor  mode  in  July  2019. We  used  the  Intermediate  Dispersion
Spectrograph (IDS) and the R632V grating together with a 1 arcsec slit
width.   The spectra  covered the  $\simeq3,500-6,000$\AA\, wavelength
range at a resolving power  of 2,500 and were reduced/calibrated using
the \textsc{pamela}/\textsc{molly} softwares.

\subsubsection{Telescopio Nazionale Galileo}

Service mode observations were carried out at the Telescopio Nazionale
Galileo at  El Roque de  los Muchachos  observatory to acquire  106 MS
high-resolution  spectra. We  used the  HARPS-N (High  Accuracy Radial
velocity   Planet   Searcher    --   North;   \citealt{Cosentino2012})
spectrograph  together  with the  1  arcsec  fibre, resulting  in  the
wavelength coverage of $\simeq$3,800$-$6,900\,\AA. The resolving power
of HARPS-N is  115,000.  In order to wavelength calibrate  the data we
used arc-lamp spectra taken with a second fibre. The data were reduced
and calibrated using the automated HARPS-N pipeline.

\subsubsection{Mercator Telescope}

We obtained  377 high-resolution MS  spectra with the  1.2\,m Mercator
telescope located  at the  El Roque de  los Muchachos  observatory. We
used  the  HERMES   spectrograph  \citep{Raskin2011},  which  provided
spectra covering the 3,770$-$9,000\AA\, range  in a single exposure at
a resolving power of 85,000. The observations were performed under the
high-resolution  (HRS)  mode  and  the data  were  reduced  using  the
dedicated  automated  data  reduction  pipeline  and  radial  velocity
toolkit        (HermesDRS)\footnote{Publicly       available        at
  http://www.mercator.iac.es/instruments/hermes/drs/}.

\subsubsection{Xinglong 2.16m Telescope}

Observations  at the  2.16  meter telescope  located  at the  Xinglong
National Station  (China) were performed  in visitor mode  during 2018
and 2019  to acquire 51 high-resolution  MS spectra. We used  the High
Resolution   fiber-fed  Spectrograph   (HRS)  \citep{Fan2016},   which
provided  echelle   spectra  covering   the  $\simeq$3,650-10,000\AA\,
wavelength range  at a  resolving power  of 49,800.  Thorium-argon arc
spectra were  taken at the  beginning and the  end of each  night. The
data   were   reduced   and   calibrated  using   the   IRAF   package
\citep{Tody1986}.

\section{Methodology and Results}
\label{s-methods}

In this section we provide details on how we derive the WD ages and MS
star metallicities, i.e. [Fe/H]  abundances. The results are presented
in  Table\,\ref{t-sdss}  for  the  SDSS  WDMS  binary  sample  and  in
Table\,\ref{t-gaia} for the \emph{Gaia} WDMS CPMPs.

\begin{table*}
\caption{\label{t-sdss}   List  of   objects,   right  ascension   and
  declination of  the 52  SDSS WDMS binaries  containing hydrogen-rich
  WDs studied in this work. For 46  of them we can derive the WD total
  ages.   We  also indicate  the  WD  effective temperatures,  surface
  gravities and masses as well as the M dwarf spectral types (Sp) that
  have  been   obtained  from  our  spectral   fitting  routine.   The
  instruments used  in the observations  are indicated too.   WD total
  ages and  MS star [Fe/H]  abundances are given  in columns 8  and 9,
  respectively.    The   WD   parameters   of   SDSSJ\,0648+3810   and
  SDSSJ\,2228+3912  are derived  fitting the  available SDSS  spectra.
  The age and WD parameters  age of SDSSJ\,0138-0016 are obtained from
  \citet{Parsons2012}.}  \setlength{\tabcolsep}{1ex} \centering
\begin{small}
\begin{tabular}{cccccccccc}
\hline
\hline
Object & ra & dec & $T_\mathrm{eff}$ (WD) & $\log$\,$g$ (WD) & Mass (WD) & Sp (MS)& Age & [Fe/H] & Instr. \\
 & (degrees) & (degrees) & (K) & (dex) & (M$_{\odot}$) &  & (Gyr) & (dex) & \\
\hline
SDSSJ\,0021-1103 &   5.49125 & -11.05878 & 10718$\pm$9    & 8.59$\pm$0.03 &  0.96$^{+0.02}_{-0.01}$  & M4 & 1.35$^{+0.09}_{-0.08}$  &   0.26$\pm$0.12 & X-Shooter\\
SDSSJ\,0054+0057 &  13.57725 &   0.96281 & 19729$\pm$42   & 7.83$\pm$0.01 & 0.543$^{+0.004}_{-0.004}$ & M4 & 4.731$^{+0.010}_{-0.003}$ &   0.22$\pm$0.12 & X-Shooter\\
SDSSJ\,0328+0017 &  52.17883 &   0.29714 & 12885$\pm$532  & 7.86$\pm$0.13 &  0.54$^{+0.06}_{-0.05}$  & M2 & 8.51$^{+3.48}_{-2.61}$  &   0.56$\pm$0.12 & X-Shooter\\
SDSSJ\,0335+0038 &  53.95242 &   0.64228 & 17993$\pm$19   & 7.92$\pm$0.01 & 0.581$^{+0.004}_{-0.004}$ & M4 & 3.62$^{+0.39}_{-0.39}$ &   0.05$\pm$0.12 & X-Shooter\\
SDSSJ\,1005+2112 & 151.26421 &  21.20111 & 20658$\pm$26   & 7.91$\pm$0.01 & 0.572$^{+0.005}_{-0.005}$ & M2 & 4.14$^{+0.1}_{-0.25}$ &   0.25$\pm$0.12 & X-Shooter\\
SDSSJ\,1114+0838 & 168.58033 &   8.64140 & 11752$\pm$59   & 8.16$\pm$0.03 &  0.70$^{+0.02}_{-0.02}$  & M5 & 0.98$^{+0.04}_{-0.04}$  &  -0.38$\pm$0.12 & X-Shooter\\
SDSSJ\,1117+0129 & 169.39633 &   1.49439 &  9018$\pm$10   & 8.25$\pm$0.03 &  0.74$^{+0.02}_{-0.02}$  & M6 & 1.60$^{+0.04}_{-0.02}$  & 0.15$\pm$0.12 & X-Shooter\\
SDSSJ\,1120+1901 & 170.05296 &  19.02412 & 13035$\pm$1263 & 7.86$\pm$0.24 &  0.53$^{+0.13}_{-0.10}$  & M3 & 8.00$^{+4.00}_{-5.60}$  &   0.17$\pm$0.12 & X-Shooter\\
SDSSJ\,1417+1301 & 214.41758  & 13.03014  & 34682$\pm$20   & 7.39$\pm$0.01 &  0.44$^{+0.01}_{-0.01}$  & M4 & - & 0.17$\pm$0.12 & X-Shooter\\
SDSSJ\,1453+0010 & 223.27405 &   0.18008 & 12025$\pm$83   & 8.32$\pm$0.03 &  0.80$^{+0.02}_{-0.02}$  & M4 & 0.86$^{+0.06}_{-0.04}$  &   0.16$\pm$0.12 & X-Shooter\\
SDSSJ\,1557+1442 & 239.33622 &  14.70491 & 12165$\pm$12   & 7.87$\pm$0.03 &  0.54$^{+0.01}_{-0.01}$  & M5 & 8.67$^{+2.14}_{-1.19}$  &  -0.17$\pm$0.12 & X-Shooter\\
SDSSJ\,1601+0505 & 240.40292 &   5.09108 & 44479$\pm$343  & 7.87$\pm$0.03 &  0.61$^{+0.01}_{-0.01}$  & M1 & 1.82$^{+0.38}_{-0.39}$  &   0.22$\pm$0.12 & X-Shooter\\
SDSSJ\,1731+0703 & 262.76886 &   7.06251 & 11888$\pm$27   & 8.60$\pm$0.03 &  0.97$^{+0.02}_{-0.02}$  & M4 & 1.04$^{+0.06}_{-0.04}$  &  -0.61$\pm$0.12 & X-Shooter\\
SDSSJ\,2044-0614 & 311.13100 &  -6.24450 & 16792$\pm$162  & 7.76$\pm$0.03 &  0.51$^{+0.01}_{-0.01}$  & M1 & -  &  -0.36$\pm$0.12 & X-Shooter\\
SDSSJ\,2312+0053 & 348.12829 &   0.88933 & 29181$\pm$39   & 8.12$\pm$0.02 &  0.70$^{+0.01}_{-0.01}$  & M2 & 0.43$^{+0.03}_{-0.03}$  &  -0.06$\pm$0.12 & X-Shooter\\
SDSSJ\,2342+1559 & 355.60696 &  15.99175 & 12885$\pm$45   & 8.06$\pm$0.03 &  0.64$^{+0.02}_{-0.02}$  & M5 & 0.99$^{+0.05}_{-0.05}$  &  -0.61$\pm$0.12 & X-Shooter\\
SDSSJ\,2350+0043 & 357.61496 &   0.73297 & 12885$\pm$2013 & 7.89$\pm$0.35 &  0.55$^{+0.20}_{-0.15}$  & M2 & 6.61$^{+5.39}_{-0.83}$  &  -0.40$\pm$0.12 & X-Shooter\\
SDSSJ\,0648+3810 & 102.05319 & 38.16831 & 20098$\pm$1341 & 7.97$\pm$0.24 &  0.61$^{+0.13}_{-0.10}$  & M1 & 1.30$^{+0.90}_{-1.29}$ &  -0.55$\pm$0.12 & SDSS/EMIR\\
SDSSJ\,1241+6007 & 190.41985 & 60.11985 & 23076$\pm$426  & 7.75$\pm$0.07 &  0.52$^{+0.02}_{-0.02}$  & M3 & - &  -0.39$\pm$0.12 & Osiris/EMIR\\
SDSSJ\,1510+4048 & 227.69042 & 40.80750 &  9573$\pm$40   & 8.45$\pm$0.07 &  0.87$^{+0.04}_{-0.05}$  & M2 & 1.50$^{+0.16}_{-0.10}$ &  -0.50$\pm$0.12 & Osiris/EMIR\\
SDSSJ\,1521+2450 & 230.31696 & 24.83787 & 36154$\pm$149  & 7.91$\pm$0.04 &  0.61$^{+0.02}_{-0.02}$  & M2 & 1.71$^{+0.55}_{-1.22}$ &   0.19$\pm$0.12 & Osiris/EMIR\\
SDSSJ\,1525+3629 & 231.32454  & 36.49589 & 9681$\pm$35    & 7.61$\pm$0.07 &  0.42$^{+0.03}_{-0.02}$  & M4 & - & -0.3$\pm$0.12 & Osiris/EMIR\\
SDSSJ\,1600+3626 & 240.24929 & 36.43608 & 23076$\pm$172  & 7.85$\pm$0.03 &  0.56$^{+0.01}_{-0.01}$  & M5 & 5.05$^{+1.66}_{-2.03}$ &  -0.12$\pm$0.12 & Osiris/EMIR\\
SDSSJ\,1605+4610 & 241.41737 & 46.17939 & 33740$\pm$280  & 7.97$\pm$0.05 &  0.64$^{+0.03}_{-0.02}$  & M0 & 0.63$^{+0.10}_{-0.12}$ &  -0.78$\pm$0.12 & Osiris/EMIR\\
SDSSJ\,1624+3217 & 246.20417 & 32.28389 & 89774$\pm$688  & 8.15$\pm$0.07 &  0.76$^{+0.04}_{-0.04}$  & M1 & 0.06$^{+0.06}_{-0.05}$ &  -0.30$\pm$0.12 & Osiris/EMIR\\
SDSSJ\,1624+3648 & 246.03358 & 36.81625 & 25012$\pm$205  & 7.99$\pm$0.03 &  0.63$^{+0.01}_{-0.01}$  & M3 & 0.73$^{+0.10}_{-0.25}$ &  -0.36$\pm$0.12 & Osiris/EMIR\\
SDSSJ\,1833+6431 & 278.37162 & 64.53104 & 54094$\pm$867  & 7.90$\pm$0.05 &  0.65$^{+0.02}_{-0.02}$  & M2 & 0.59$^{+0.08}_{-0.09}$ &  -0.51$\pm$0.12 & Osiris/EMIR\\
SDSSJ\,1834+4137 & 278.72116 & 41.63269 &  9456$\pm$24   & 8.12$\pm$0.05 &  0.67$^{+0.03}_{-0.03}$  & M4 & 1.42$^{+0.03}_{-0.09}$ &  -0.85$\pm$0.12 & Osiris/EMIR\\
SDSSJ\,2228+3912 & 337.09474 & 39.21106 & 25794$\pm$1745 & 7.81$\pm$0.24 &  0.55$^{+0.12}_{-0.08}$  & M3 & 7.23$^{+4.71}_{-1.46}$ &   0.06$\pm$0.12 & SDSS/EMIR\\
SDSSJ\,0003-0503*&   0.98723 & -5.05909 & 19967$\pm$131  & 8.07$\pm$0.02 &  0.66$^{+0.01}_{-0.01}$   & M4 & 0.72$^{+0.11}_{-0.11}$ &  0.05$\pm$0.12 & X-Shooter\\
SDSSJ\,0005-0544*&   1.49948 & -5.73780 & 32748$\pm$224  & 7.73$\pm$0.03 &  0.53$^{+0.01}_{-0.01}$   & M2 & 8.27$^{+1.07}_{-0.28}$ &  0.06$\pm$0.12 & X-Shooter\\ 
SDSSJ\,0036+0700*&   9.01079 &  7.01311 & 36105$\pm$49   & 7.87$\pm$0.02 & 0.589$^{+0.009}_{-0.009}$ & M4 & 3.89$^{+0.94}_{-0.95}$ &  0.30$\pm$0.12 & X-Shooter\\ 
SDSSJ\,0052-0051*&  13.03508 & -0.85961 & 11933$\pm$94   & 8.02$\pm$0.03 &  0.61$^{+0.02}_{-0.02}$   & M4 & 1.53$^{+0.23}_{-0.52}$ &  0.06$\pm$0.12 & X-Shooter\\
SDSSJ\,0111+0009* & 17.84954 & 0.15981 &  12326$\pm$47  & 7.76$\pm$0.03 & 0.50$^{+0.01}_{-0.01}$   & M2  & - & -0.46$\pm$0.12 & X-Shooter\\
SDSSJ\,0138-0016*&  24.71475 & -0.27267 &  3570$\pm$110  & 7.92$\pm$0.02 &  0.54$^{+0.01}_{-0.01}$   & M5 & 9.50$^{+0.30}_{-0.20}$ & -0.56$\pm$0.12 & X-Shooter\\
SDSSJ\,0256-0730*&  44.04421 & -7.50683 & 10194$\pm$68   & 8.84$\pm$0.04 &  1.09$^{+0.02}_{-0.02}$   & M5 & 1.99$^{+0.02}_{-0.02}$ & -0.27$\pm$0.12 & X-Shooter\\
SDSSJ\,0258+0109*&  44.57446 &  1.16278 & 36873$\pm$123  & 7.75$\pm$0.02 & 0.540$^{+0.007}_{-0.007}$ & M3 & 5.87$^{+0.05}_{-0.39}$ &  0.23$\pm$0.12 & X-Shooter\\ 
SDSSJ\,0321-0016*&  50.40225 & -0.27511 & 31096$\pm$32   & 7.88$\pm$0.02 & 0.584$^{+0.009}_{-0.008}$ & M5 & 2.95$^{+0.88}_{-1.13}$ &  0.17$\pm$0.12 & X-Shooter\\
SDSSJ\,0325-0111*&  51.29517 & -1.18725 & 10499$\pm$14   & 8.13$\pm$0.04 &  0.68$^{+0.02}_{-0.02}$   & M2 & 1.19$^{+0.03}_{-0.04}$ & -0.36$\pm$0.12 & X-Shooter\\
SDSSJ\,0331-0054*&  52.88383 & -0.91483 & 30742$\pm$30   & 7.96$\pm$0.01 & 0.622$^{+0.005}_{-0.005}$ & M3 & 1.07$^{+0.08}_{-0.08}$ &  0.09$\pm$0.12 & X-Shooter\\
SDSSJ\,0824+1723*& 126.12092 & 17.39594 & 12476$\pm$52   & 7.86$\pm$0.03 &  0.54$^{+0.01}_{-0.01}$   & M3 & 9.40$^{+2.33}_{-0.47}$ & -0.10$\pm$0.12 & X-Shooter\\
SDSSJ\,0832-0430*& 128.23002 & -4.51285 & 16064$\pm$85   & 8.01$\pm$0.01 & 0.623$^{+0.006}_{-0.005}$ & M1 & 0.85$^{+0.02}_{-0.03}$ & -0.76$\pm$0.12 & X-Shooter\\
SDSSJ\,0916-0031*& 139.00617 & -0.52494 & 19130$\pm$63   & 8.30$\pm$0.02 &  0.79$^{+0.01}_{-0.01}$   & M4 & 0.46$^{+0.05}_{-0.01}$ &  0.30$\pm$0.12 & X-Shooter\\
SDSSJ\,0933+0926*& 143.29962  & 9.44508 & 30401$\pm$18   & 7.63$\pm$0.01 &  0.50$^{+0.002}_{-0.002}$ & M5 & - &-0.07$\pm$0.12 & X-Shooter\\ 
SDSSJ\,1023+0427*& 155.89271 &  4.45617 & 20498$\pm$68   & 7.89$\pm$0.02 &  0.56$^{+0.01}_{-0.01}$   & M4 & 4.21$^{+0.42}_{-0.21}$ &  0.18$\pm$0.12 & X-Shooter\\ 
SDSSJ\,1040+0834*& 160.23950 &  8.57267 & 10254$\pm$8    & 8.00$\pm$0.03 &  0.60$^{+0.02}_{-0.02}$   & M5 & 1.55$^{+0.23}_{-0.31}$ & -0.09$\pm$0.12 & X-Shooter\\
SDSSJ\,1405+0409*& 211.39554 &  4.15183 & 20716$\pm$88   & 8.15$\pm$0.02 &  0.71$^{+0.02}_{-0.01}$   & M4 & 0.51$^{+0.03}_{-0.03}$ & -0.30$\pm$0.12 & X-Shooter\\
SDSSJ\,1527+1007*& 231.93379 & 10.12289 & 34079$\pm$52   & 7.86$\pm$0.02 & 0.588$^{+0.009}_{-0.009}$ & M3 & 1.31$^{+0.24}_{-1.30}$ & -0.17$\pm$0.12 & X-Shooter\\
SDSSJ\,1539+0922*& 234.89431 &  9.37265 & 11183$\pm$143  & 8.72$\pm$0.03 &  1.04$^{+0.02}_{-0.02}$   & M5 & 1.53$^{+0.02}_{-0.01}$ &  0.29$\pm$0.12 & X-Shooter\\ 
SDSSJ\,1558+0231*& 239.72171 &  2.52731 & 30062$\pm$9    & 7.79$\pm$0.01 & 0.548$^{+0.004}_{-0.003}$ & M4 & 7.24$^{+0.38}_{-0.29}$ &  0.07$\pm$0.12 & X-Shooter\\ 
SDSSJ\,1624-0022*& 246.13192 & -0.38006 & 26291$\pm$203  & 7.92$\pm$0.03 &  0.60$^{+0.01}_{-0.01}$   & M3 & 1.40$^{+0.30}_{-0.45}$ &  0.02$\pm$0.12 & X-Shooter\\ 
SDSSJ\,2341-0947*& 355.49262 & -9.78794 &  9433$\pm$26   & 8.27$\pm$0.04 &  0.76$^{+0.02}_{-0.02}$   & M4 & 1.41$^{+0.02}_{-0.02}$ &  0.07$\pm$0.12 & X-Shooter\\ 
\hline
\end{tabular} 
\end{small}
\begin{minipage}{\textwidth}
Those systems that are reanalysed from \citet{Rebassa16AMR} are indicated by * after their names.
\end{minipage}
\end{table*}

\subsection{M dwarf metallicities of the SDSS WDMS binary sample}
\label{s-sdssmetal}

We  obtained  the  M  dwarf [Fe/H]  abundances  from  their  $K$-band,
near-infrared  X-Shooter  and  EMIR spectra  following  the  procedure
described in \citet{Newton2014}. Note that at these wavelengths the WD
contributions are negligible  (see Figure\,\ref{f-spec}).  This method
uses  a semi-empirical  multivariate  linear regression  based on  the
equivalent width of  the 2,205/2,209 nm Na~{\sc  i} absorption doublet
to yield  [Fe/H] values with an  accuracy of 0.12 dex.   To derive the
equivalent width values we  corrected the systemic (radial) velocities
and we normalized  the fluxes in the 2,194--2,220  nm wavelength range
fitting a  third-order spline function.   In this process  we excluded
the  Na~{\sc  i}  doublet  absorption   feature.   We  then  used  the
trapezoidal  rule to  integrate the  flux within  the 2,204--2,210  nm
region of the absorption doublet.

\subsection{WD ages of the SDSS WDMS binary sample}
\label{sdss-param}

We first run a decomposition/fitting  routine to the optical X-Shooter
and OSIRIS spectra  of the binaries to determine the  spectral type of
the M dwarfs  and subtract their flux  contribution \citep[left panels
  of  Figure\,\ref{f-spec},  for   details  see][]{Rebassa2007}.   The
normalised Balmer  lines of the  residual WD spectra were  then fitted
with a grid of hydrogen-rich  WD model atmosphere spectra \citep[][and
  unpublished improvements]{Koester2010}  in order to  measure $T_{\rm
  eff}$ and $\log{g}$ (see  the right panels of Figure\,\ref{f-spec}).
When  appropriate, we  accounted for  the 3D  corrections provided  by
\protect\citet{Tremblay2013} and, thus, we linearly interpolated these
values in  the cooling sequences developed  by the La Plata  Group for
three   different  metallicities   ($Z$=0.001,  \citealt{Althaus2015};
$Z$=0.01, \citealt{Renedo2010}; and $Z$=0.02, \citealt{Camisassa2016})
to  obtain the  WD masses  and  total ages,  i.e. WD  cooling plus  MS
progenitor  lifetime\footnote{The  WD  masses  and  cooling  ages  are
  practically identical in the three  cases, however the MS progenitor
  lifetimes significantly  vary as function of  metallicity. Note also
  that   for    ultra-massive   WDs    we   used   the    sequence   of
  \citet{Camisassa2019}  for   the  three  metallicities.}.    The  WD
evolutionary sequences  for $Z$=0.001  and $Z$=0.02 were  derived from
the  full evolutionary  history of  their progenitor  stars, from  the
Zero-Age-MS all  the way to the  WD phase.  These WD  models take into
account all the relevant energy  sources that govern the WD evolution,
including  the  energy released  by  the  crystallization process,  as
latent  heat  and as  gravitational  energy  by the  phase  separation
process.   The progenitor  lifetimes adopted  in these  sequences were
interpolated from Table\,2 of  \citet{Miller2016} and they adopted the
initial-to-final mass relation  of the same work, which  is similar to
the    semi-empirical    relations    found    in    \citet{Catalan08,
  Cummings2018}. The sequences of \citet{Renedo2010} for $Z$=0.01 take
into account all the relevant  processes involved in the WD evolution,
however they  do not  provide the  progenitor lifetimes.   We obtained
those interpolating from Table\,2 of \citet{Miller2016}.

Given that we know the [Fe/H] abundances (Section\,\ref{s-sdssmetal}),
and hence the  $Z$ values\footnote{We fitted a  fifth order polynomial
  between   [Fe/H]   and    $Z$   conversions,   calculated   assuming
  \citet{Asplund2005}      solar     abundances,      resulting     in
  $Z=0.01230+0.02560\times[\mathrm{Fe/H}]+0.03291\times[\mathrm{Fe/H}]^{2}+0.02833\times[\mathrm{Fe/H}]^{3}+0.01284\times[\mathrm{Fe/H}]^{4}+0.00223\times[\mathrm{Fe/H}]^{5}$.},
of the MS companions (assumed to  be the same for the WD progenitors),
we interpolated the measured $Z$ between the three $Z$/total-age pairs
obtained for each system from  the evolutionary WD sequences to derive
the corresponding  total ages.  In the  cases where  the metallicities
were larger than  0.02 (the highest value provided by  the models), we
linearly  extrapolated   the  ages  from  the   cooling  sequences  of
\citet{Camisassa2016}.  Taking  into account  the errors  in effective
temperature and  surface gravity, we  calculated the total  age errors
following the same approach.

It has to be  emphasised that in this work we used  a more updated set
of  cooling sequences  as well  as a  different initial-to-final  mass
relation than in \citet{Rebassa16AMR}. Therefore, we have recalculated
the total ages of the binaries  studied in that work and included them
in     Table\,\ref{t-sdss}.     Moreover,      as     indicated     in
Section\,\ref{s-gtc}, two SDSS WDMS binaries were only observed by the
GTC  using the  EMIR spectrograph.  In these  two cases,  in order  to
obtain  the total  ages, we  used  the WD  effective temperatures  and
surface gravities reported by \citet{Rebassa2016b}, which were derived
by fitting  the optical SDSS  publicly available spectra of  these two
binaries.

Of   the   52   SDSS   WDMS    binaries   analysed   here   (23   from
\citealt{Rebassa16AMR}),            6           have            masses
$\la0.50$\,M$_\mathrm{\odot}$. Such low-mass WDs can not be formed via
single stellar evolution because  their progenitors have main sequence
lifetimes  longer than  the Hubble  time. Therefore,  a binary  origin
involving mass transfer interactions  is generally required to explain
these objects  \citep[e.g.][]{Rebassa2011}. Because of this  reason we
can not  derive the ages  of these 6  systems.  A possibility  is that
these unresolved  SDSS WDMS binaries  are in fact close  binaries that
evolved through  a common  envelope phase  but have  low inclinations,
thus implying no radial velocity  variation detection. This issue will
be further discussed in Section\,\ref{s-discussion}.

\subsection{MS metallicities of the \emph{Gaia} WDMS CPMP sample}
\label{s-metgaia}

We obtained  534 high-resolution  spectra (Table\,\ref{t-log})  of 349
unique  companion stars  in our  sample\footnote{Several targets  were
  observed more than  once at the same  and/or different telescopes.},
26 of which have spectral types  earlier than $\simeq$F0, 260 are F, G
or K-type stars, 6 are giants  of luminosity class between III and IV,
and 55  are M  dwarfs (45  of spectral type  M3 or  earlier and  10 of
spectral types  M4-M5).  The  lack of near-infrared  spectroscopy does
not allow measuring the [Fe/H] abundances  of the M dwarfs in the same
way as we  have performed for the SDSS sample.  Hence, we attempted to
measure the abundances of the 311 F,  G, K and early M (M3 or earlier)
as well as giant companions via detailed spectroscopic fits.

We   determined  the   effective   temperatures,  surface   gravities,
microturbulent   velocities   and    [Fe/H]   abundances   using   the
\textsc{tgvit}  code \citep{takeda05}.   This  routine implements  the
iron ionisation plus the iron  equilibrium conditions as well as match
of  the curve  of growth.   Such a  methodology is  widely applied  to
solar-type stars of spectral types between F5 and K2/K3. We used a set
of  well-defined 302  Fe~{\sc  i}  and 28  Fe~{\sc  ii}  lines in  the
analysis.   The  code  makes  use  of  ATLAS9,  plane-parallel,  local
thermodynamic  equilibrium (LTE)  atmosphere models  \citep{kurucz93}.
Uncertainties in the stellar parameters are statistical, that is, each
stellar parameter is progressively changed from the converged solution
until a  value in  which any  of the  aforementioned conditions  is no
longer fulfilled.

For low-mass  stars, M  dwarfs, we  determined the  stellar parameters
using          the           procedures          developed          by
\cite{Maldonado2015}\footnote{https://github.com/jesusmaldonadoprado/mdslines}. In
brief, this  routine uses  as a temperature  diagnostic the  ratios of
pseudo-equivalent widths of spectral  features, while calibrations for
the stellar  metallicity are derived  from combinations and  ratios of
features.  The  temperatures and  metallicities thus derived  are used
together  with  photometric estimates  of  surface  gravity, mass  and
radius to calibrate empirical relations for these parameters.

The procedures  described above converged  to reliable fits in  235 of
the cases. These companions stars are illustrated as black solid dots,
as  well as  their  WD primaries,  in  Figure\,\ref{fig:hr}.  For  the
remaining  76   objects  the  available   spectra  were  of   too  low
signal-to-noise ratio ($\la$30), the number  of measured Fe lines were
too low, or there were problems in measuring and correcting the radial
velocity for  a proper  determination of  the [Fe/H]  abundances.  For
those objects with  more than one [Fe/H]  determination from available
multiple spectra, we used the average  values.  In the few cases where
the individual  [Fe/H] values  considerably disagreed, we  adopted the
value corresponding to the spectrum with higher signal-to-noise ratio.

\subsection{WD ages of the \emph{Gaia} WDMS CPMP sample}
\label{s-agegaia}

Low-resolution optical  spectra were obtained  for 214 WDs at  the WHT
and  INT telescopes  (Table\,\ref{t-log}),  however only  67 of  their
companions     have    available     high-resolution    MS     spectra
(Section\,\ref{s-metgaia}). Of these 67 WDs, 15 are hydrogen-rich (DA)
WDs  that unfortunately  have  too noisy  spectra  for measurement  of
reliable  stellar  parameters, 14  are  featureless  (DC) WDs,  3  are
helium-rich  (DB) WDs,  1 is  a  metal-rich (DZ)  WD and  only 34  are
hydrogen-rich (DA) WDs  with good fits to their  spectra. This implies
we can only attempt  to determine ages for 34 of  the 235 binaries for
which we have obtained a metallicity value (Section\,\ref{s-metgaia}).
As  a consequence,  rather than  measuring the  WD stellar  parameters
(hence ages) from their available  good-quality spectra, we decided to
derive them using the  corresponding \emph{Gaia} EDR3 \citep{Gaia2020,
  Riello2020}   photometry  and   parallaxes\footnote{Note  that   the
  \emph{Gaia}   WDMS    CPMPs   were    selected   using    DR2   data
  (Figure\,\ref{fig:hr}). However, we used  the newest EDR3 to measure
  the WD ages.}.  Compared to using spectroscopy alone, this holds the
potential of  providing a factor  $\simeq$7 increase in the  number of
age-metallicity pairs.

\begin{figure}
    \centering
    \includegraphics[width=\linewidth]{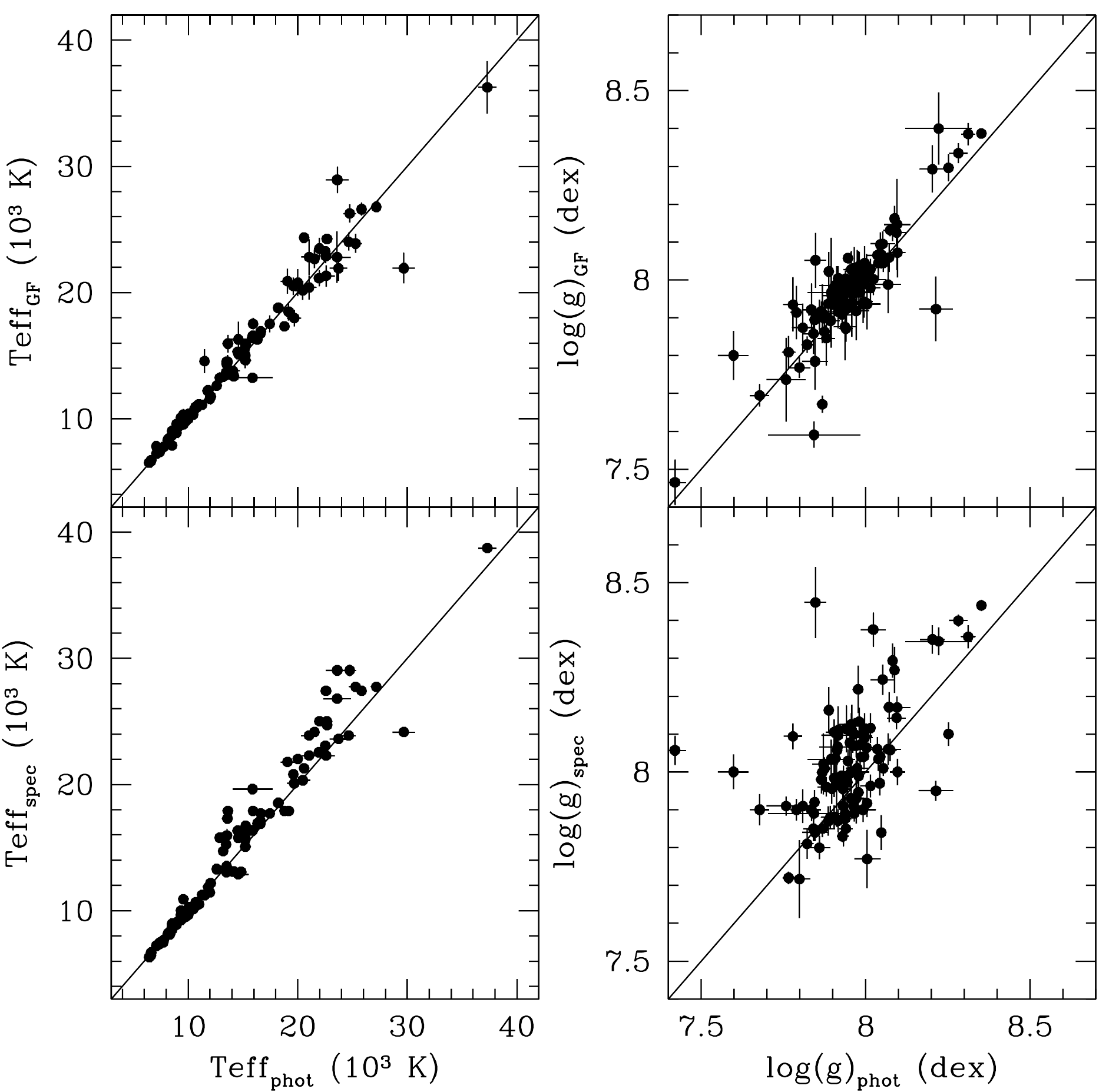}
    \caption{Top  panels:  a  comparison   between  the  WD  effective
      temperatures (left)  and surface  gravities (right)  measured in
      this  work from  \emph{Gaia} EDR3  photometry and  parallaxes as
      compared to  those obtained by  \citet{gentile-fusillo2019} from
      \emph{Gaia} DR2 data. Bottom panels:  the same but comparing the
      photometric  EDR3 values  to those  obtained from  our follow-up
      spectroscopy.}
    \label{fig:comp}
\end{figure}

\begin{table*}
\caption{\label{t-gaia}  Object  names,  WD  and  MS  coordinates,  WD
  effective temperatures,  surface gravities and masses  measured from
  \emph{Gaia}  photometry  and   spectroscopy  (when  available),  the
  corresponding total ages and the  MS star [Fe/H] abundances (assumed
  to be  same as for the  WD progenitors) of the  235 \emph{Gaia} WDMS
  binaries analysed in this work. For 46 WDs the masses are too low to
  derive an age.  The complete table can be found in the supplementary
  material, where  the \emph{Gaia}  source IDs  of each  component are
  also provided.}  \setlength{\tabcolsep}{0.4ex} \centering
\begin{small}
\begin{tabular}{cccccccccccc}
\hline
\hline
   Object     &    WD ra/dec          &   MS ra/dec         & $T_\mathrm{eff-phot}$ & log\,$g_\mathrm{phot}$ & Mass$_\mathrm{phot}$ & Age$_\mathrm{phot}$ & $T_\mathrm{eff-spec}$ & log\,$g_\mathrm{spec}$ & Mass$_\mathrm{spec}$ & Age$_\mathrm{spec}$ & [Fe/H]\\
               & (degrees) & (degress) & (K) & (dex) &
(M$_{\odot}$) & (Gyr) & (K) & (dex) & (M$_{\odot}$) & (Gyr) & (dex) \\
\hline
J0012+2328    &    2.96163/23.48678   &    3.00612/23.47178 &      8109$\pm$210   &   8.17$\pm$0.06  &   0.70$^{+0.04}_{-0.04}$ &  1.77$^{+0.10}_{-0.08}$ &      -         &         -       &       -      &     -	     & -0.18$\pm$0.03  \\
J0021+2531    &    5.31779/25.52627   &    5.31696/25.52431 &      9332$\pm$126   &   8.49$\pm$0.03  &   0.90$^{+0.02}_{-0.02}$ &  1.57$^{+0.02}_{-0.01}$ &      -         &         -       &       -      &     -	     &  0.16$\pm$0.05  \\
J0033+4443    &    8.26320/44.73679   &    8.25721/44.72989 &     10198$\pm$164   &   7.98$\pm$0.03  &   0.59$^{+0.02}_{-0.02}$ &  1.99$^{+0.48}_{-1.83}$ &  10141$\pm$63  &   8.07$\pm$0.06 &     0.64$^{+0.03}_{-0.04}$  &   1.35$^{+0.12}_{-0.29}$  & -0.03$\pm$0.02  \\
J0045+1421    &   11.34235/14.34570   &   11.33858/14.36269 &      4886$\pm$34    &   7.81$\pm$0.02  &   0.49$^{+0.01}_{-0.01}$ &     -                &      -         &         -       &       -      &     -	     & -0.26$\pm$0.03   \\
J0048+1333    &   12.18230/13.55794   &   12.18096/13.55589 &     12588$\pm$1543  &   8.45$\pm$0.14  &   0.88$^{+0.09}_{-0.09}$ &  0.65$^{+0.06}_{-0.38}$ &      -         &         -       &       -      &     -	     & 0.27$\pm$0.5  \\
J0055+3321    &   13.81336/33.35159   &   13.81654/33.35214 &     18190$\pm$1274  &   7.90$\pm$0.08  &   0.55$^{+0.04}_{-0.04}$ &  6.50$^{+2.60}_{-5.80}$ &      -         &         -       &       -      &     -	     & 0.42$\pm$0.04  \\
J0103+6108    &   15.79464/61.12871   &   15.79292/61.13567 &      9317$\pm$270   &   8.02$\pm$0.06  &   0.61$^{+0.03}_{-0.03}$ &  1.48$^{+0.14}_{-1.36}$ &      -         &         -       &       -      &     -	     & -0.10$\pm$0.03 \\
J0108+7018    &   17.12171/70.30084   &   17.13725/70.30042 &      5351$\pm$154   &   7.93$\pm$0.09  &   0.55$^{+0.05}_{-0.04}$ & 10.21$^{+1.79}_{-2.83}$ &      -         &         -       &       -      &     -	     & -0.11$\pm$0.04  \\
J0112+0454    &   18.03588/ 4.91914   &   18.03363/ 4.91625 &     19160$\pm$744   &   8.07$\pm$0.04  &   0.66$^{+0.02}_{-0.02}$ &  0.68$^{+0.13}_{-0.15}$ &  20098$\pm$211 &   8.06$\pm$0.04 &    0.65$^{+0.02}_{-0.02}$  &   0.69$^{+0.13}_{-0.13}$ & -0.03$\pm$0.04  \\
J0115+1534    &   18.98383/15.58056   &   18.98204/15.58008 &     24039$\pm$577   &   7.95$\pm$0.02  &   0.59$^{+0.01}_{-0.01}$ &  5.65$^{+1.20}_{-1.03}$ &      -         &         -       &       -      &     -	     & 0.63$\pm$0.13 \\
J0119+6218    &   19.83586/62.30148   &   19.83471/62.30539 &     16166$\pm$336   &   7.94$\pm$0.03  &   0.57$^{+0.02}_{-0.01}$ &  7.55$^{+2.21}_{-0.14}$ &      -         &         -       &       -      &     -	     & 0.41$\pm$0.03 \\
\hline
\end{tabular} 
\end{small}
\end{table*}

All the WDs that are part of our \emph{Gaia} WDMS CPMPs are associated
to measured  parallaxes and  $G$, $G_\mathrm{BP}$  and $G_\mathrm{RP}$
magnitudes from \emph{Gaia} EDR3.  This allowed for derivations of the
corresponding  effective  temperatures  and surface  gravities,  hence
masses and ages, by interpolating the observed $G$ absolute magnitudes
and  $G_\mathrm{BP}-G_\mathrm{RP}$ colours  in  the cooling  sequences
developed by  the La  Plata group  for pure-hydrogen  atmospheres (see
Section\,\ref{sdss-param}).   The synthetic  $G$, $G_\mathrm{BP}$  and
$G_\mathrm{RP}$ absolute magnitudes were incorporated  by us to the WD
cooling  sequences  integrating  the  flux  of  the  associated  model
atmosphere  spectra \citep{Koester2010}  over  the corresponding  EDR3
pass-bands.   It  has to  be  emphasised  that before  performing  the
interpolation  the  observed  EDR3   magnitudes  were  corrected  from
extinction using the 3D  maps of \citet{Lallement2014, Capitanio2017}.
As expected from  their close distance, the  extinction correction was
rather low in most cases  with an average value of $A_\mathrm{g}$=0.04
mag  and  a  maximum  correction  of  $A_\mathrm{g}$=0.26  mag,  where
$A_\mathrm{g}$  was  obtained from  $A_\mathrm{V}$  using  the law  of
\citet{Fitzpatrick2019}   and    assuming   $R_\mathrm{V}$=3.1.    The
effective  temperature and  surface  gravity errors  were obtained  by
propagating the parallax and photometric errors.

\begin{figure*}
    \centering
    \includegraphics[width=\columnwidth]{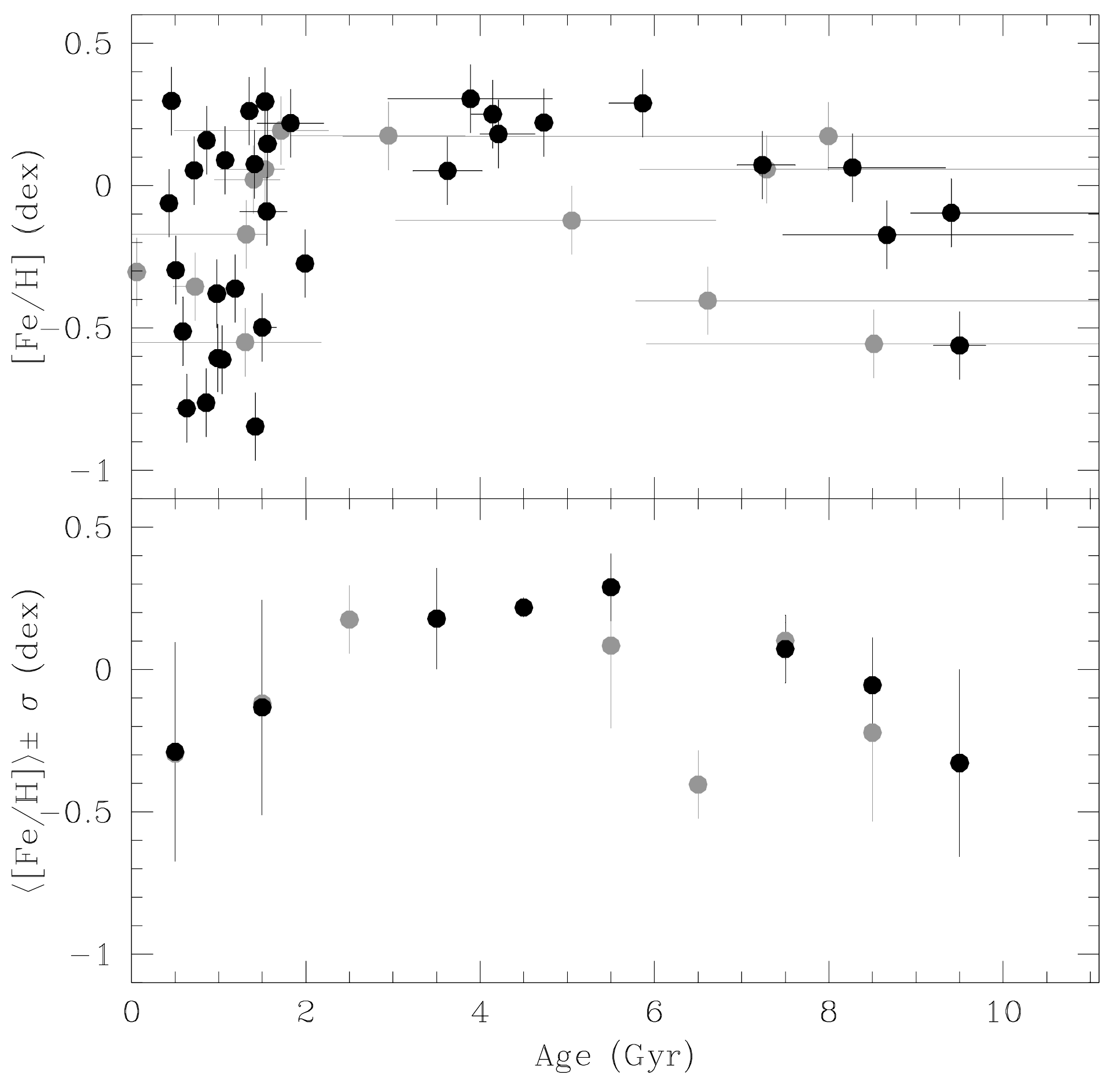}
    \includegraphics[width=\columnwidth]{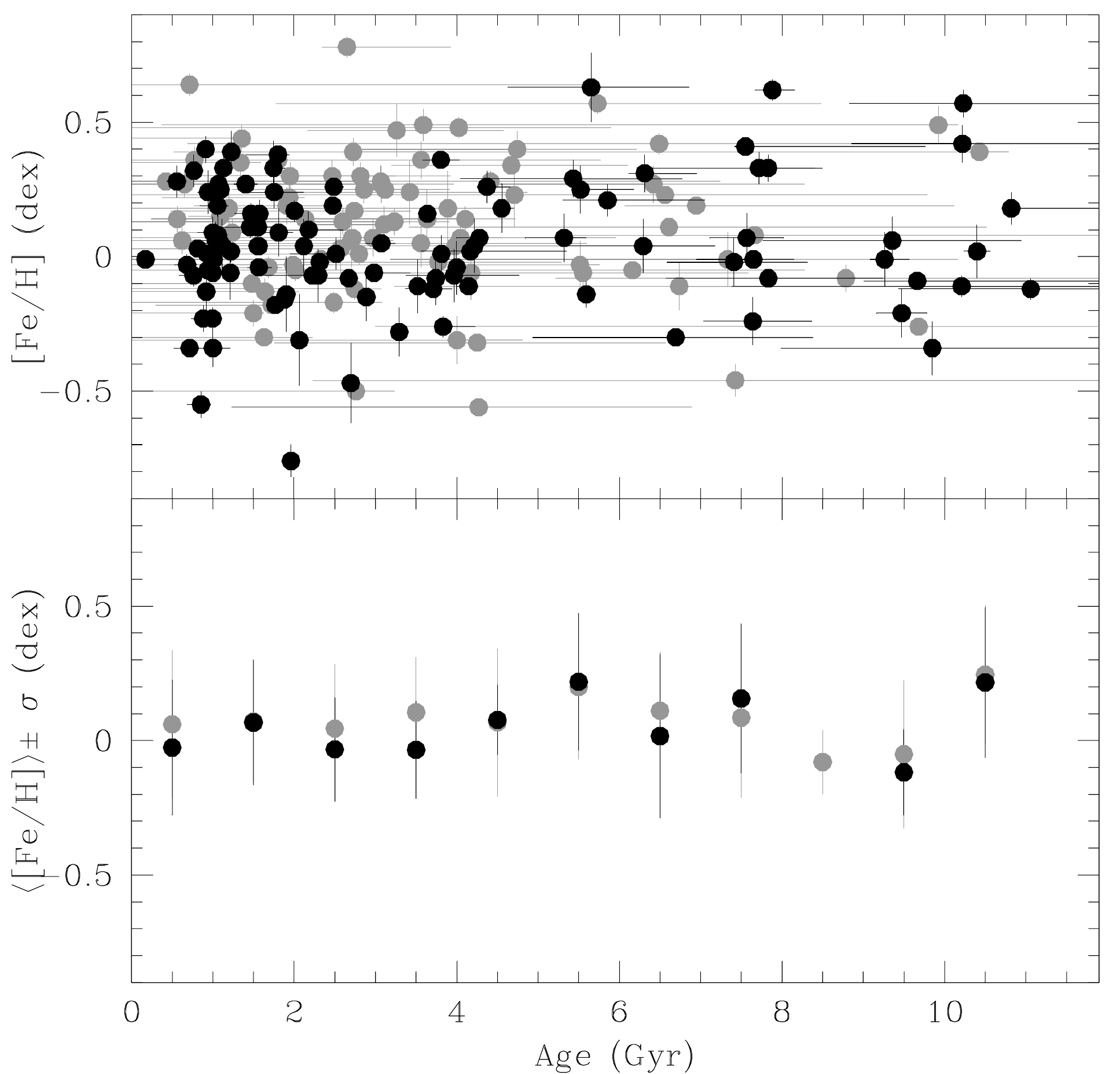}
    \caption{Top  left panel:  the  age-metallicity relation  obtained
      from SDSS WDMS binaries. Black dots represent ages with relative
      errors less than 30 per  cent. Bottom left panel: average [Fe/H]
      values for 1  Gyr bins, together with  their standard deviations
      for  the  entire sample  (grey)  and  for  the sample  with  age
      relative errors  below 30  per cent  (black). Right  panels: the
      same but  for \emph{Gaia} WDMS  CPMPs.}
    \label{fig:amr_sdss_gaia}
\end{figure*} 

It  is important  to note  that we  find a  relatively large  fraction
($\simeq$19  per  cent)   of  low-mass  ($\la$0.5\,M$_\mathrm{\odot}$)
WDs. This is  intriguing since these are in resolved  CPMP systems and
their  progenitors  should  have  evolved  like  single  stars,  hence
producing higher mass WDs.  Half  of the these presumably low-mass WDs
have photometric effective  temperatures under $\simeq$6,000\,K, hence
they could be non-DA WDs for which the age we me measure is uncertain.
A systematic shift towards low  masses below $\simeq$6,000\,K has also
been reported by \citet{McCleery2020}. For the hotter WDs, pure-helium
DB  models  would   yield  even  lower  masses.    Hence,  a  possible
explanation is that the atmospheres of these WDs are not hydrogen- nor
helium-dominated. Another  possibility is  that these objects  are, or
were, triple  systems formed by an  inner close binary (for  example a
double degenerate)  and an  outer MS  companion.  We  will investigate
these hypotheses in a separate publication.

An obvious question  is whether the stellar  parameters measured using
both methods  (spectroscopy and photometry) are  consistent.  To check
this we  derived the effective  temperatures and surface  gravities of
107 hydrogen-rich DA  WDs that we observed  at the WHT and  INT in the
same way  as described in Section\,\ref{sdss-param}  (without the need
of subtracting the MS companion contribution, since these are resolved
WDs  in  CPMPs).   Hereafter, we  consider  $T_\mathrm{eff-spec}$  and
log($g$)$_\mathrm{spec}$ as the measured values from spectroscopy. The
photometric        parameters,        $T_\mathrm{eff-phot}$        and
log\,$g_\mathrm{phot}$,  were derived  as  described  in the  previous
paragraphs from  the available  \emph{Gaia} EDR3  data. In  the bottom
panels  of Figure\,\ref{fig:comp}  we compare  the stellar  parameters
thus obtained. If we define $\tau$ as
\begin{equation}
\tau = \frac{|\mathrm{spec}\,\mathrm{value} - \mathrm{phot}\,\mathrm{value}|}{\sqrt{\mathrm{spec}\,\mathrm{\sigma^2}+\mathrm{phot}\,\mathrm{\sigma^2}}},
\end{equation}
where  `value'  indicates  either  effective  temperature  of  surface
gravity  and  $\sigma$ the  corresponding  errors,  we find  that  the
$T_\mathrm{eff}$  and  log\,$g$  measurements  are  consistent  within
2$\tau$  for  $\simeq$30  and  $\simeq$50   per  cent  of  the  cases,
respectively.   \citet{Tremblay2020}   argued  that   the  photometric
effective temperatures  they measured for  a sample of  89 \emph{Gaia}
DR2 hydrogen-rich WDs  within 40\,pc were 2.7  per cent underestimated
as compared to their  spectroscopic measurements.  By considering this
effect,  the  percentage  of  WDs with  consistent  spectroscopic  and
photometric  effective  temperatures  in  our  sample  increases  from
$\simeq$30 to $\simeq$45 per cent. The percentages of agreement within
2$\tau$   further   increase  to   $\simeq$70   per   cent  for   both
$T_\mathrm{eff}$ and  log\,$g$ if  we systematically add  a 0.015\,dex
uncertainty to the  log\,$g$ measurements and a  150\,K uncertainty to
the effective temperature values. This  clearly reveals the effect the
small errors  in our  measurements have in  quantifying the  degree of
consistency  between  the  two  methods. We  thus  conclude  that  the
spectroscopic and photometric measurements  are broadly consistent. We
will further discuss this issue in Section\,\ref{s-discussion}.

For completeness,  we show in  the top panels  of Fig.\,\ref{fig:comp}
the comparison between our photometric  values and those obtained from
\emph{Gaia} DR2 data by \citet{gentile-fusillo2019}. Visual inspection
suggests  the two  sets  of data  are in  good  agreement. Indeed,  in
$\simeq$75/85 per cent of  the cases the effective temperature/surface
gravity measurements are consistent within 2$\tau$.

\begin{figure*}
    \centering
    \includegraphics[width=\columnwidth]{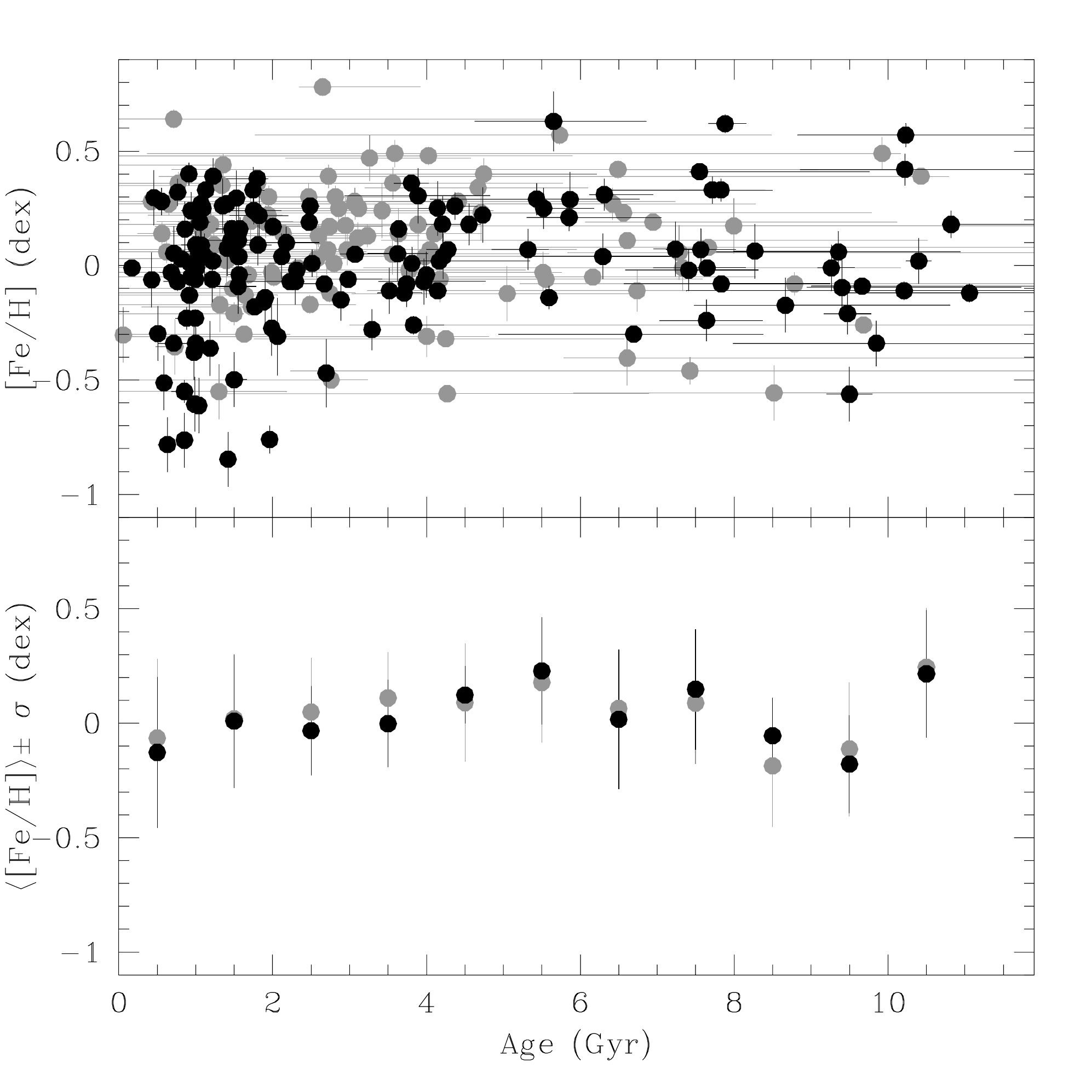}
    \includegraphics[width=\columnwidth]{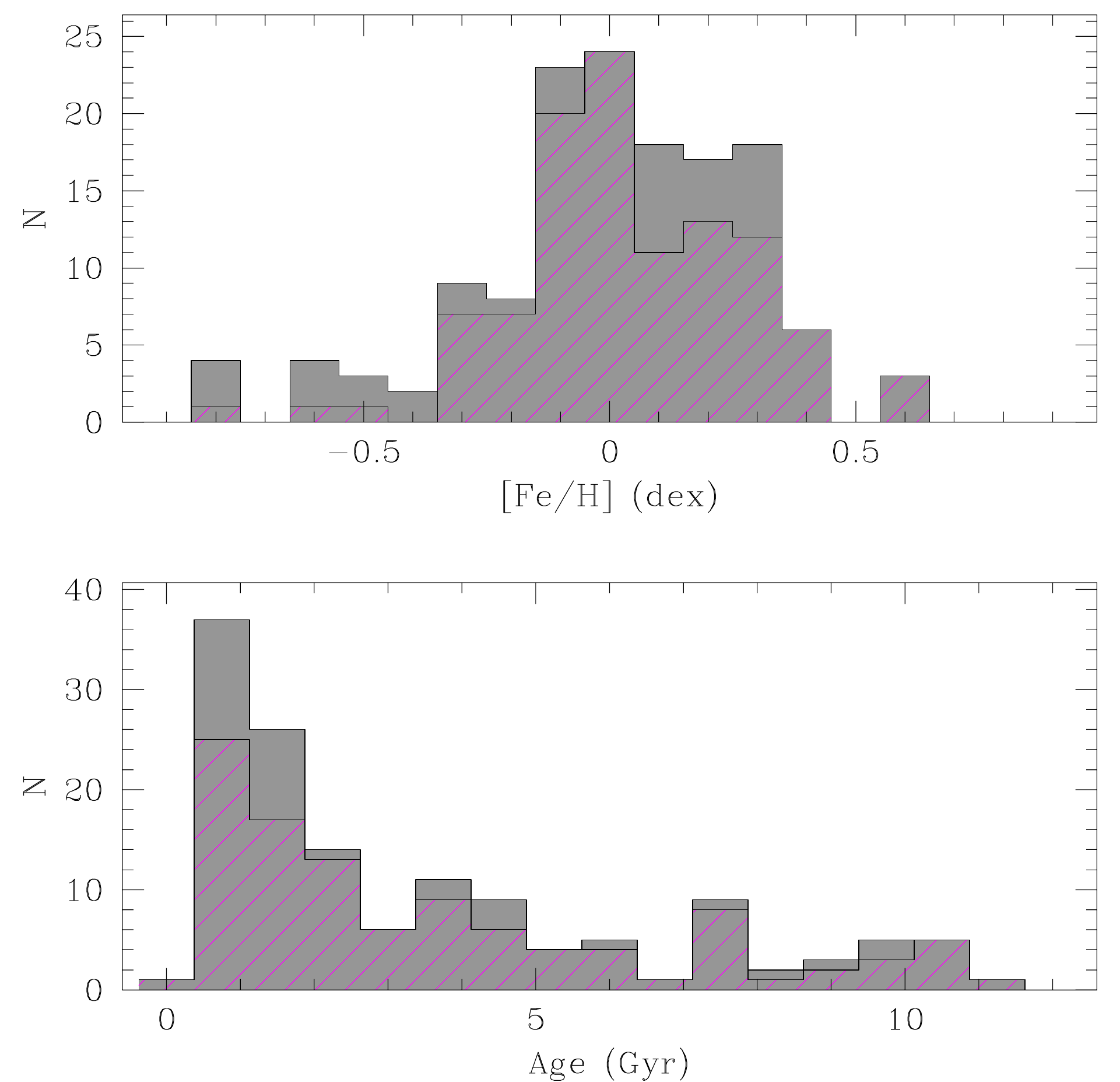}
    \caption{Left panels: the  same as Figure\,\ref{fig:amr_sdss_gaia}
      but for  the entire sample  studied in  this work (46  SDSS WDMS
      binaries  plus  189  \emph{Gaia}   WDMS  CPMPs).   There  is  no
      correlation between age and  metallicity.  Right panels: the age
      (bottom;  relative  error less  than  30  per cent)  and  [Fe/H]
      abundance  (top)  corresponding distributions.  For  comparative
      purposes  the age  and  [Fe/H] abundance  distributions for  the
      \emph{Gaia} WDMS binary sample are shown in magenta.}
    \label{fig:amr_full}
\end{figure*}

\section{Discussion: the age-metallicity relation}
\label{s-discussion}

The AMR that results from the SDSS  WDMS binary sample is shown in the
top  left  panel  of Figure\,\ref{fig:amr_sdss_gaia},  where  a  clear
scatter of  [Fe/H] abundances ($>$0.2  dex) is observed at  most ages.
This is  corroborated in  the bottom  left panel  of the  same figure,
where we  show the average  [Fe/H] per 1 Gyr  bin. It is  worth noting
that in  $\simeq$20 per  cent of  the cases  the age  measurements are
subjected  to substantial  uncertainties ($>$30  per cent  of relative
error; grey  points in Figure\,\ref{fig:amr_sdss_gaia}).  This  is due
to the  relatively large uncertainties  in the measured  WD parameters
that translate into  larger age uncertainties and, mainly,  due to the
fact that  the determined WD masses  are low ($\la$0.55\,M$_{\odot}$).
The lower the WD mass, the longer its progenitor spent on the MS. This
MS lifetime is  very sensitive to the mass of  the WD progenitor star,
thus even small errors in the  derived WD masses translate into rather
different ages  spent on  the MS  and, as  a consequence,  into larger
total age uncertainties.  Given that low-mass WDs evolve from low-mass
MS stars  that require a longer  time to leave the  main sequence, the
percentage   of  ages   with  larger   uncertainties  is   higher  for
intermediate and old ages ($\ga$5 Gyr).

It  is also  important  to  mention that  these  unresolved SDSS  WDMS
binaries are expected to have  evolved avoiding mass transfer episodes
based on the lack of radial velocity variations. However, as discussed
in Section\,\ref{sdss-param},  it is possible  that a small  number of
the SDSS WDMS  systems are in fact post common  envelope binaries with
low orbital inclinations (preventing  the detection of radial velocity
variations).  For  those  systems,  the measured  ages  would  not  be
accurate.  The probability of non-detection of a close WDMS binary due
to a  low-inclination is  $\simeq$15 per cent  \citep{Nebot2011}. This
could  explain why  the  three  most metal-poor  stars  in the  sample
([Fe/H]$<-$0.7 dex) appear to be younger than 2 Gyrs.

The \emph{Gaia} WDMS CPMP AMR is illustrated in the top right panel of
Figure\,\ref{fig:amr_sdss_gaia} and  the corresponding  average [Fe/H]
per 1 Gyr  bin in the bottom right  panel. In the same way  as we have
observed for the SDSS sample, the scatter of [Fe/H] abundances becomes
apparent at  all ages. It has  to be noted  that the WD ages  for this
sample were  obtained from  the available \emph{Gaia}  EDR3 photometry
and that the WD stellar parameters  obtained from these data are found
to be  broadly consistent with  those obtained from  the spectroscopic
fits (Section\,\ref{s-agegaia}).  To investigate  whether or  not this
issue affects the  result obtained, we derived  the (spectroscopic) WD
ages  of the  34 DA  WDs  for which  the stellar  parameters are  also
measured  from spectroscopy  (Section\,\ref{s-agegaia}). These  values
are  included in  Table\,\ref{t-gaia}. We  found the  same scatter  of
[Fe/H] despite  the fact that  the individual ages vary,  as expected,
for some objects.  Therefore, we can safely conclude  that the scatter
of metallicities in the \emph{Gaia} WDMS CPMP AMR is real.

It is  also important to  emphasise that  the WD cooling  sequences we
have adopted to  determine the stellar parameters  are those developed
for hydrogen-rich DA  WDs. The lack of spectroscopy does  not allow us
to  confirm  this  assumption  for those  WDs  with  only  \emph{Gaia}
photometry available. Thus, we expect a  fraction of non-DA WDs in our
sample. From our spectroscopic sub-sample  of 67 objects, we calculate
fractions  of  $\simeq$73  per  cent  DAs,  $\simeq$4  per  cent  DBs,
$\simeq$1  per cent  DZs and  $\simeq$20  per cent  DCs, although  the
latter   (visual)  classification   is   likely   biased  because   of
signal-to-noise ratio issues of the spectra.  Assuming these fractions
are the same for the entire \emph{Gaia} WDMS CPMP sample, we reach the
following  conclusions:  (i)  $\simeq4$  per  cent  of  our  ages  are
underestimated because we are  using pure-hydrogen sequences to derive
the  ages of  DB WDs.   This  is because  photometric masses  obtained
assuming  pure-helium  DB sequences  are  generally  lower than  those
obtained using pure-hydrogen  models \citep{gentile-fusillo2019}. This
implies lower MS progenitor masses  and larger progenitor lifetimes as
compared to those obtained using pure-hydrogen models; (ii) given that
$\simeq$40   per   cent   of   WDs   have   He-rich   atmospheres   at
$\simeq5,000-7000$\,K  \citep{McCleery2020},  we expect  approximately
half of the spectroscopically (visually)  classified objects as DC WDs
to  have   He-dominated  atmospheres.   This  implies   an  additional
$\simeq$10 per cent of underestimated  ages in the total sample; (iii)
$\simeq$1 per cent  of the WDs are expected to  have other atmospheric
composition apart from hydrogen and helium,  such as DZ WDs.  In these
cases the derived ages are uncertain. In summary, we expect $\simeq$14
per cent  of the \emph{Gaia} WDMS  CPMP ages to be  underestimated and
$\simeq$1 per cent to be uncertain.

In the top left panel of Fig.\,\ref{fig:amr_full} we show the AMR that
results from combining  our SDSS and the full  \emph{Gaia} WDMS binary
samples,  a  total  of  235  age-[Fe/H]  pairs.   The  average  [Fe/H]
abundances per  Gyr bin and the  standard deviations are shown  in the
left   bottom   panel   of   the   same   figure   and   provided   in
Table\,\ref{t:std}. The  sample is dominated by  young objects ($\la$5
Gyr; see bottom right panel  of Fig.\,\ref{fig:amr_full}), but it also
samples  both  intermediate and  old  ages.   Unfortunately, very  few
objects in our total sample have  ages larger than 10\,Gyr, which does
not allow us to study the properties  of the AMR relation for very old
ages in a robust way.  The drop of systems with ages $>$10\,Gyr can be
understood as follows.  In order to achieve an age longer than 10 Gyrs
a WD  requires one of  the following properties:  (1) a very  low mass
($\simeq$0.5\,M$_{\odot}$)  thus implying  a very  long main  sequence
progenitor lifetime  and a short  cooling age (hence a  relatively hot
tempreature;  $>$10,000  K); (2)  a  very  cool effective  temperature
($\leq$4000  K),  thus  implying  a  long  cooling  age  and  a  short
progenitor  lifetime (hence  a  relatively more  massive white  dwarf;
$\ga$0.6\,M$_{\odot}$); (3) a low mass ($\simeq$0.55\,M$_{\odot}$) and
a  low   effective  temperature  ($\sim$6000  K).    As  explained  in
Section\,\ref{s-sample}, the  SDSS WDMS sample suffers  from selection
effects against cool  WDs and therefore the  probability for detecting
very old  age systems decreases.   For the \emph{Gaia}  population, we
are   analysing   objects   with    average   distances   of   150\,pc
(Section\,\ref{s-sample}),  which   defines  a  sample  close   to  be
volume-limited.   In   a  volume-limited  sample  not   only  hot  WDs
($>$10,000   K)  are   intrinsically  less   numerous  but   also  the
completeness of  cold ($<$5000 K)  and less luminous WDs  decreases to
$\simeq$60  per cent  \citep{Torres2021}.  These  effects imply  that,
overall, it becomes intrinsically more difficult to find WDs with ages
longer than 10 Gyrs.

Like \citet{Casagrande2011},  we find no steep  decline in metallicity
for ages in  the $\simeq$9-10 Gyr range, a trend  that was reported by
\citet{Bergemann14}.    The  AMR   we   obtained  fluctuates   between
approximately  -0.5  and +0.5\,dex  with  a  peak centered  at  around
[Fe/H]=0 (see the top right panel of Fig.\,\ref{fig:amr_full}) and, as
expected from the  previous discussion, it does not  show any apparent
correlation.   This  result  is  in agreement  with  several  previous
studies that  analysed the  AMR in  which the  ages were  derived from
Str\"omgren     photometry     \citep{Feltzing2001,     Nordstrom2004,
  Casagrande2011},   spectroscopy   and  luminosities   obtained   via
parallaxes  \citep{Buder2019, DelgadoMena2019,  Nissen2020}, isochrone
matching   \citep{Hayden2020},   open   clusters   \citep{Carraro1998,
  Pancino2010}    and     asteroseismology    \citep{SilvaAguirre2018,
  Miglio2021}.  The analysis performed in  this work using a different
technique  for measuring  stellar  ages reaches  the same  conclusion.
Self-enrichment     of     gas     in     star     forming     regions
\citep{Pilyugin+Edmunds1996}  or episodic  gas in-fall  onto the  disc
\citep{Koppen+Hensler2005} are proposed physical mechanisms to explain
the  lack of  correlation between  age  and metallicity  in the  solar
neighbourhood.   However, the  most accepted  scenario invokes  radial
migration, a physical mechanism in which metal-rich stars that form in
the  inner   disc  migrate   to  the   outer  and   metal-poorer  disc
\citep{SellwoodBinney2002,   Roskar2008,   Minchev2011,   Minchev2013,
  Feuillet2019}.  The  validity of this scenario  is further supported
by  the  fact  that  our   AMR  displays  young  ($\simeq$1\,Gyr)  but
metal-rich    ([Fe/H]$>$2     dex)    objects    (left     panel    of
Fig.\,\ref{fig:amr_full}).   It is  widely  accepted  that such  stars
cannot form in the solar neighbourhood and, as a consequece, must have
migrated  from  a  more  metal-rich enviroment  in  the  inner  Galaxy
\citep{Haywood2008, Brunetti2011}.

\begin{table}
\caption{\label{t:std} $\langle$[Fe/H]$\rangle  \pm \sigma$  for 1~Gyr
  bins  obtained from  the full  WDMS  binary sample  studied in  this
  work. Only the  age values with relative uncertainties  under 30 per
  cent have been  considered (black solid dots in the  bottom panel of
  Fig.\,\ref{fig:amr_full}).  For completeness,  we  also provide  the
  values     for     the     \emph{Gaia}    and     SDSS     samples.}
\setlength{\tabcolsep}{3ex} \centering
\begin{tabular}{cccc}
\hline
\hline
& Full sample &  \emph{Gaia} sample  & SDSS sample  \\
\hline
  Age    &    $\langle$[Fe/H]$\rangle \pm \sigma$ &   $\langle$[Fe/H]$\rangle \pm \sigma$ &  $\langle$[Fe/H]$\rangle \pm \sigma$ \\
  (Gyr)  &    (dex) &  (dex) & (dex)\\
\hline
 0.5 &  -0.13 $\pm$  0.33 & -0.03 $\pm$ 0.25 &  -0.29 $\pm$  0.39 \\
 1.5 &   0.01 $\pm$  0.29 &  0.07 $\pm$ 0.23 &  -0.13 $\pm$  0.38 \\
 2.5 &  -0.03 $\pm$  0.19 & -0.03 $\pm$ 0.19 &     -              \\
 3.5 &   0.00 $\pm$  0.19 & -0.03 $\pm$ 0.18 &   0.18 $\pm$  0.18 \\
 4.5 &   0.12 $\pm$  0.13 &  0.08 $\pm$ 0.13 &   0.22 $\pm$  0.03 \\
 5.5 &   0.23 $\pm$  0.23 &  0.22 $\pm$ 0.26 &   0.29 $\pm$  0.12 \\
 6.5 &   0.02 $\pm$  0.31 &  0.02 $\pm$ 0.31 &    -               \\
 7.5 &   0.15 $\pm$  0.26 &  0.16 $\pm$ 0.28 &   0.07 $\pm$  0.12 \\
 8.5 &  -0.06 $\pm$  0.17 &   -              &  -0.06 $\pm$  0.17 \\
 9.5 &  -0.18 $\pm$  0.21 & -0.12 $\pm$ 0.16 &  -0.33 $\pm$  0.33 \\
10.5 &   0.22 $\pm$  0.28 &  0.22 $\pm$ 0.28 &     -              \\
\hline
\end{tabular} 
\end{table}

\section{Conclusions}
\label{s-concl}

Dating  a star  is a  challenging  task. For  decades, this  difficult
endeavour  has   been  pursued   for  single  stars   using  different
techniques, which  has allowed analysing the  observational properties
of  the age-metallicity  relation in  the solar  neighbourhood.  These
studies  seem  to converge  to  the  same  result,  i.e. there  is  no
correlation  between  age and  metallicity  as  traced by  the  [Fe/H]
abundances. In this work we have used an alternative way for measuring
stellar  ages  based on  the  analysis  of white  dwarf-main  sequence
binaries from which  we can obtain accurate white dwarf  ages and main
sequence [Fe/H] abundances.  The  age-metallicity relation that we have
obtained  displays a  scatter of  approximately $\pm$  0.5 dex  at all
ages.  This  is  yet  another  robust  confirmation  of  the  lack  of
correlation between age and metallicity in the solar neighbourhood.

\section*{Acknowledgments}

Project  supported  by  a  2019 Leonardo  Grant  for  Researchers  and
Cultural  Creators,  BBVA  Foundation.    The  Foundation  accepts  no
responsibility for  the opinions, statements and  contents included in
the  project  and/or  the  results thereof,  which  are  entirely  the
responsibility of the authors.

ARM acknowledges additional support from  the MINECO under the Ram\'on
y  Cajal program  (RYC-2016-20254). JM  acknowledges support  from the
Accordo Attuativo  ASI-INAF n. 2018.22.HH.O, Partecipazione  alla fase
B1 della missione Ariel. RR has received funding from the postdoctoral
fellowship programme  Beatriu de Pin\'os,  funded by the  Secretary of
Universities and Research (Government of Catalonia) and by the Horizon
2020 programme of research and  innovation of the European Union under
the  Maria   Sk\l{}odowska-Curie  grant   agreement  No   801370.   ST
acknowledges support from the  MINECO under the AYA2017-86274-P grant,
and the AGAUR grant SGR-661/2017. MJH  was supported by the UK Science
and Technology  Facilities Council studentship ST/R505195/1.   BTG was
supported by  a Leverhulme Research  Fellowship and the UK  STFC grant
ST/T000406/1. PET, TC  and MH have received funding  from the European
Research Council under the European  Union's Horizon 2020 research and
innovation programme n. 677706  (WD3D). 

Based on  observations made  with the  Gran Telescopio  Canarias (GTC;
programmes GTC6-16B,  GTC1-17A, GTC26-17B and GTC6-18A),  installed in
the Spanish Observatorio  del Roque de los Muchachos  of the Instituto
de Astrof\'isica de Canarias, in the  island of La Palma. This work is
based on observations made with ESO Telescopes at the La Silla Paranal
Observatory under programme ID 0101.B-0130. Based on observations made
with the  Telescopio Nazionale Galileo and  Mercator Telescope awarded
to the  International Time  Programme ITP18\_8. Based  on observations
made with the William Herschel  Telescope (programmes C12 and C26) and
Isaac Newton  Telescope (programme P4W)  operated on the island  of La
Palma  by  the  Isaac  Newton  Group  of  Telescopes  in  the  Spanish
Observatorio  del  Roque   de  los  Muchachos  of   the  Instituto  de
Astrof\'isica de Canarias. We acknowledge  the support of the staff of
the Xinglong 2.16  m telescope.  This work was  partially supported by
the Open Project  Program of the Key Laboratory  of Optical Astronomy,
National Astronomical Observatories, Chinese Academy of Sciences.

MJH acknowledges Saskia Prins, the  support astronomer at Mercator for
her invaluable help.

This work  has made use of  data from the European  Space Agency (ESA)
mission {\it  Gaia} (\url{https://www.cosmos.esa.int/gaia}), processed
by  the {\it  Gaia}  Data Processing  and  Analysis Consortium  (DPAC,
\url{https://www.cosmos.esa.int/web/gaia/dpac/consortium}).    Funding
for the DPAC has been provided by national institutions, in particular
the  institutions   participating  in  the  {\it   Gaia}  Multilateral
Agreement.

\section*{Data Availability}

The   data   underlying   this    article   are   available   in   the
manuscript.  Supplementary  material  will  be  shared  on  reasonable
request to the corresponding author.




\begin{thebibliography}{}
\makeatletter
\relax
\def\mn@urlcharsother{\let\do\@makeother \do\$\do\&\do\#\do\^\do\_\do\%\do\~}
\def\mn@doi{\begingroup\mn@urlcharsother \@ifnextchar [ {\mn@doi@}
  {\mn@doi@[]}}
\def\mn@doi@[#1]#2{\def\@tempa{#1}\ifx\@tempa\@empty \href
  {http://dx.doi.org/#2} {doi:#2}\else \href {http://dx.doi.org/#2} {#1}\fi
  \endgroup}
\def\mn@eprint#1#2{\mn@eprint@#1:#2::\@nil}
\def\mn@eprint@arXiv#1{\href {http://arxiv.org/abs/#1} {{\tt arXiv:#1}}}
\def\mn@eprint@dblp#1{\href {http://dblp.uni-trier.de/rec/bibtex/#1.xml}
  {dblp:#1}}
\def\mn@eprint@#1:#2:#3:#4\@nil{\def\@tempa {#1}\def\@tempb {#2}\def\@tempc
  {#3}\ifx \@tempc \@empty \let \@tempc \@tempb \let \@tempb \@tempa \fi \ifx
  \@tempb \@empty \def\@tempb {arXiv}\fi \@ifundefined
  {mn@eprint@\@tempb}{\@tempb:\@tempc}{\expandafter \expandafter \csname
  mn@eprint@\@tempb\endcsname \expandafter{\@tempc}}}

\bibitem[\protect\citeauthoryear{{Althaus}, {C{\'o}rsico}, {Isern}  \&
  {Garc{\'\i}a-Berro}}{{Althaus} et~al.}{2010}]{Althaus2010}
{Althaus} L.~G.,  {C{\'o}rsico} A.~H.,  {Isern} J.,   {Garc{\'\i}a-Berro} E.,
  2010, \mn@doi [\aapr] {10.1007/s00159-010-0033-1}, \href
  {https://ui.adsabs.harvard.edu/abs/2010A&ARv..18..471A} {18, 471}

\bibitem[\protect\citeauthoryear{{Althaus}, {Camisassa}, {Miller Bertolami},
  {C{\'o}rsico}  \& {Garc{\'\i}a-Berro}}{{Althaus} et~al.}{2015}]{Althaus2015}
{Althaus} L.~G.,  {Camisassa} M.~E.,  {Miller Bertolami} M.~M.,  {C{\'o}rsico}
  A.~H.,   {Garc{\'\i}a-Berro} E.,  2015, \mn@doi [\aap]
  {10.1051/0004-6361/201424922}, \href
  {https://ui.adsabs.harvard.edu/abs/2015A&A...576A...9A} {576, A9}

\bibitem[\protect\citeauthoryear{{Andrews}, {Chanam{\'e}}  \&
  {Ag{\"u}eros}}{{Andrews} et~al.}{2017}]{andrews2017}
{Andrews} J.~J.,  {Chanam{\'e}} J.,   {Ag{\"u}eros} M.~A.,  2017, \mn@doi
  [\mnras] {10.1093/mnras/stx2000}, \href
  {https://ui.adsabs.harvard.edu/abs/2017MNRAS.472..675A} {472, 675}

\bibitem[\protect\citeauthoryear{{Asplund}, {Grevesse}  \& {Sauval}}{{Asplund}
  et~al.}{2005}]{Asplund2005}
{Asplund} M.,  {Grevesse} N.,   {Sauval} A.~J.,  2005, in {Barnes} III T.~G.,
  {Bash} F.~N.,  eds,  Astronomical Society of the Pacific Conference Series
  Vol. 336, Cosmic Abundances as Records of Stellar Evolution and
  Nucleosynthesis. p.~25

\bibitem[\protect\citeauthoryear{{Barrientos} \& {Chanam{\'e}}}{{Barrientos} \&
  {Chanam{\'e}}}{2021}]{Barrientos2021}
{Barrientos} M.,  {Chanam{\'e}} J.,  2021, arXiv e-prints, \href
  {https://ui.adsabs.harvard.edu/abs/2021arXiv210207790B} {p. arXiv:2102.07790}

\bibitem[\protect\citeauthoryear{{Barry}}{{Barry}}{1988}]{Barry88}
{Barry} D.~C.,  1988, \mn@doi [\apj] {10.1086/166848}, \href
  {https://ui.adsabs.harvard.edu/abs/1988ApJ...334..436B} {334, 436}

\bibitem[\protect\citeauthoryear{{Bensby}, {Feltzing}  \& {Oey}}{{Bensby}
  et~al.}{2014}]{Bensby2014}
{Bensby} T.,  {Feltzing} S.,   {Oey} M.~S.,  2014, \mn@doi [\aap]
  {10.1051/0004-6361/201322631}, \href
  {http://adsabs.harvard.edu/abs/2014A26A...562A..71B} {562, A71}

\bibitem[\protect\citeauthoryear{{Bergemann} et~al.,}{{Bergemann}
  et~al.}{2014}]{Bergemann14}
{Bergemann} M.,  et~al., 2014, \mn@doi [\aap] {10.1051/0004-6361/201423456},
  \href {https://ui.adsabs.harvard.edu/abs/2014A&A...565A..89B} {565, A89}

\bibitem[\protect\citeauthoryear{{Brunetti}, {Chiappini}  \&
  {Pfenniger}}{{Brunetti} et~al.}{2011}]{Brunetti2011}
{Brunetti} M.,  {Chiappini} C.,   {Pfenniger} D.,  2011, \mn@doi [\aap]
  {10.1051/0004-6361/201117566}, \href
  {https://ui.adsabs.harvard.edu/abs/2011A&A...534A..75B} {534, A75}

\bibitem[\protect\citeauthoryear{{Buder} et~al.,}{{Buder}
  et~al.}{2019}]{Buder2019}
{Buder} S.,  et~al., 2019, \mn@doi [\aap] {10.1051/0004-6361/201833218}, \href
  {https://ui.adsabs.harvard.edu/abs/2019A&A...624A..19B} {624, A19}

\bibitem[\protect\citeauthoryear{{Camisassa}, {Althaus}, {C{\'o}rsico},
  {Vinyoles}, {Serenelli}, {Isern}, {Miller Bertolami}  \&
  {Garc{\'\i}a-Berro}}{{Camisassa} et~al.}{2016}]{Camisassa2016}
{Camisassa} M.~E.,  {Althaus} L.~G.,  {C{\'o}rsico} A.~H.,  {Vinyoles} N.,
  {Serenelli} A.~M.,  {Isern} J.,  {Miller Bertolami} M.~M.,
  {Garc{\'\i}a-Berro} E.,  2016, \mn@doi [\apj] {10.3847/0004-637X/823/2/158},
  \href {https://ui.adsabs.harvard.edu/abs/2016ApJ...823..158C} {823, 158}

\bibitem[\protect\citeauthoryear{{Camisassa} et~al.,}{{Camisassa}
  et~al.}{2019}]{Camisassa2019}
{Camisassa} M.~E.,  et~al., 2019, \mn@doi [\aap] {10.1051/0004-6361/201833822},
  \href {https://ui.adsabs.harvard.edu/abs/2019A&A...625A..87C} {625, A87}

\bibitem[\protect\citeauthoryear{{Capitanio}, {Lallement}, {Vergely},
  {Elyajouri}  \& {Monreal-Ibero}}{{Capitanio} et~al.}{2017}]{Capitanio2017}
{Capitanio} L.,  {Lallement} R.,  {Vergely} J.~L.,  {Elyajouri} M.,
  {Monreal-Ibero} A.,  2017, \mn@doi [\aap] {10.1051/0004-6361/201730831},
  \href {https://ui.adsabs.harvard.edu/abs/2017A&A...606A..65C} {606, A65}

\bibitem[\protect\citeauthoryear{{Carraro}, {Ng}  \& {Portinari}}{{Carraro}
  et~al.}{1998}]{Carraro1998}
{Carraro} G.,  {Ng} Y.~K.,   {Portinari} L.,  1998, \mn@doi [\mnras]
  {10.1046/j.1365-8711.1998.01460.x}, \href
  {https://ui.adsabs.harvard.edu/abs/1998MNRAS.296.1045C} {296, 1045}

\bibitem[\protect\citeauthoryear{{Casagrande}, {Sch{\"o}nrich}, {Asplund},
  {Cassisi}, {Ram{\'\i}rez}, {Mel{\'e}ndez}, {Bensby}  \&
  {Feltzing}}{{Casagrande} et~al.}{2011}]{Casagrande2011}
{Casagrande} L.,  {Sch{\"o}nrich} R.,  {Asplund} M.,  {Cassisi} S.,
  {Ram{\'\i}rez} I.,  {Mel{\'e}ndez} J.,  {Bensby} T.,   {Feltzing} S.,  2011,
  \mn@doi [\aap] {10.1051/0004-6361/201016276}, \href
  {https://ui.adsabs.harvard.edu/abs/2011A&A...530A.138C} {530, A138}

\bibitem[\protect\citeauthoryear{{Casagrande} et~al.,}{{Casagrande}
  et~al.}{2016}]{Casagrande2016}
{Casagrande} L.,  et~al., 2016, \mn@doi [\mnras] {10.1093/mnras/stv2320}, \href
  {https://ui.adsabs.harvard.edu/abs/2016MNRAS.455..987C} {455, 987}

\bibitem[\protect\citeauthoryear{{Catal{\'a}n}, {Isern}, {Garc{\'\i}a-Berro}
  \& {Ribas}}{{Catal{\'a}n} et~al.}{2008}]{Catalan08}
{Catal{\'a}n} S.,  {Isern} J.,  {Garc{\'\i}a-Berro} E.,   {Ribas} I.,  2008,
  \mn@doi [\mnras] {10.1111/j.1365-2966.2008.13356.x}, \href
  {https://ui.adsabs.harvard.edu/abs/2008MNRAS.387.1693C} {387, 1693}

\bibitem[\protect\citeauthoryear{{Cosentino} et~al.,}{{Cosentino}
  et~al.}{2012}]{Cosentino2012}
{Cosentino} R.,  et~al., 2012, in {McLean} I.~S.,  {Ramsay} S.~K.,   {Takami}
  H.,  eds,  Society of Photo-Optical Instrumentation Engineers (SPIE)
  Conference Series Vol. 8446, Ground-based and Airborne Instrumentation for
  Astronomy IV. p. 84461V, \mn@doi{10.1117/12.925738}

\bibitem[\protect\citeauthoryear{{Cummings}, {Kalirai}, {Tremblay},
  {Ramirez-Ruiz}  \& {Choi}}{{Cummings} et~al.}{2018}]{Cummings2018}
{Cummings} J.~D.,  {Kalirai} J.~S.,  {Tremblay} P.~E.,  {Ramirez-Ruiz} E.,
  {Choi} J.,  2018, \mn@doi [\apj] {10.3847/1538-4357/aadfd6}, \href
  {https://ui.adsabs.harvard.edu/abs/2018ApJ...866...21C} {866, 21}

\bibitem[\protect\citeauthoryear{{Delgado Mena} et~al.,}{{Delgado Mena}
  et~al.}{2019}]{DelgadoMena2019}
{Delgado Mena} E.,  et~al., 2019, \mn@doi [\aap] {10.1051/0004-6361/201834783},
  \href {https://ui.adsabs.harvard.edu/abs/2019A&A...624A..78D} {624, A78}

\bibitem[\protect\citeauthoryear{{Edvardsson}, {Andersen}, {Gustafsson},
  {Lambert}, {Nissen}  \& {Tomkin}}{{Edvardsson} et~al.}{1993}]{Edvardsson93}
{Edvardsson} B.,  {Andersen} J.,  {Gustafsson} B.,  {Lambert} D.~L.,  {Nissen}
  P.~E.,   {Tomkin} J.,  1993, \aap, \href
  {http://adsabs.harvard.edu/abs/1993A26A...275..101E} {275, 101}

\bibitem[\protect\citeauthoryear{{Eisenstein} et~al.,}{{Eisenstein}
  et~al.}{2011}]{Eisenstein2011}
{Eisenstein} D.~J.,  et~al., 2011, \mn@doi [\aj] {10.1088/0004-6256/142/3/72},
  \href {https://ui.adsabs.harvard.edu/abs/2011AJ....142...72E} {142, 72}

\bibitem[\protect\citeauthoryear{{El-Badry} \& {Rix}}{{El-Badry} \&
  {Rix}}{2018}]{el-badry2018}
{El-Badry} K.,  {Rix} H.-W.,  2018, \mn@doi [\mnras] {10.1093/mnras/sty2186},
  \href {https://ui.adsabs.harvard.edu/abs/2018MNRAS.480.4884E} {480, 4884}

\bibitem[\protect\citeauthoryear{{El-Badry}, {Rix}  \& {Heintz}}{{El-Badry}
  et~al.}{2021}]{el-badry2021}
{El-Badry} K.,  {Rix} H.-W.,   {Heintz} T.~M.,  2021, \mn@doi [\mnras]
  {10.1093/mnras/stab323}, \href
  {https://ui.adsabs.harvard.edu/abs/2021MNRAS.tmp..394E} {}

\bibitem[\protect\citeauthoryear{{Fan} et~al.,}{{Fan} et~al.}{2016}]{Fan2016}
{Fan} Z.,  et~al., 2016, \mn@doi [\pasp] {10.1088/1538-3873/128/969/115005},
  \href {https://ui.adsabs.harvard.edu/abs/2016PASP..128k5005F} {128, 115005}

\bibitem[\protect\citeauthoryear{{Feltzing} \& {Chiba}}{{Feltzing} \&
  {Chiba}}{2013}]{Feltzing13}
{Feltzing} S.,  {Chiba} M.,  2013, \mn@doi [\nar]
  {10.1016/j.newar.2013.06.001}, \href
  {https://ui.adsabs.harvard.edu/abs/2013NewAR..57...80F} {57, 80}

\bibitem[\protect\citeauthoryear{{Feltzing}, {Holmberg}  \&
  {Hurley}}{{Feltzing} et~al.}{2001}]{Feltzing2001}
{Feltzing} S.,  {Holmberg} J.,   {Hurley} J.~R.,  2001, \mn@doi [\aap]
  {10.1051/0004-6361:20011119}, \href
  {https://ui.adsabs.harvard.edu/abs/2001A&A...377..911F} {377, 911}

\bibitem[\protect\citeauthoryear{{Feuillet}, {Frankel}, {Lind}, {Frinchaboy},
  {Garc{\'\i}a-Hern{\'a}ndez}, {Lane}, {Nitschelm}  \&
  {Roman-Lopes}}{{Feuillet} et~al.}{2019}]{Feuillet2019}
{Feuillet} D.~K.,  {Frankel} N.,  {Lind} K.,  {Frinchaboy} P.~M.,
  {Garc{\'\i}a-Hern{\'a}ndez} D.~A.,  {Lane} R.~R.,  {Nitschelm} C.,
  {Roman-Lopes} A.,  2019, \mn@doi [\mnras] {10.1093/mnras/stz2221}, \href
  {https://ui.adsabs.harvard.edu/abs/2019MNRAS.489.1742F} {489, 1742}

\bibitem[\protect\citeauthoryear{{Fitzpatrick}, {Massa}, {Gordon}, {Bohlin}  \&
  {Clayton}}{{Fitzpatrick} et~al.}{2019}]{Fitzpatrick2019}
{Fitzpatrick} E.~L.,  {Massa} D.,  {Gordon} K.~D.,  {Bohlin} R.,   {Clayton}
  G.~C.,  2019, \mn@doi [\apj] {10.3847/1538-4357/ab4c3a}, \href
  {https://ui.adsabs.harvard.edu/abs/2019ApJ...886..108F} {886, 108}

\bibitem[\protect\citeauthoryear{{Fouesneau}, {Rix}, {von Hippel}, {Hogg}  \&
  {Tian}}{{Fouesneau} et~al.}{2019}]{Fouesneau2019}
{Fouesneau} M.,  {Rix} H.-W.,  {von Hippel} T.,  {Hogg} D.~W.,   {Tian} H.,
  2019, \mn@doi [\apj] {10.3847/1538-4357/aaee74}, \href
  {https://ui.adsabs.harvard.edu/abs/2019ApJ...870....9F} {870, 9}

\bibitem[\protect\citeauthoryear{{Gaia Collaboration} et~al.,}{{Gaia
  Collaboration} et~al.}{2016}]{Gaia2016}
{Gaia Collaboration} et~al., 2016, \mn@doi [\aap]
  {10.1051/0004-6361/201629272}, \href
  {https://ui.adsabs.harvard.edu/abs/2016A&A...595A...1G} {595, A1}

\bibitem[\protect\citeauthoryear{{Gaia Collaboration} et~al.,}{{Gaia
  Collaboration} et~al.}{2018a}]{Gaia2018}
{Gaia Collaboration} et~al., 2018a, \mn@doi [\aap]
  {10.1051/0004-6361/201833051}, \href
  {https://ui.adsabs.harvard.edu/abs/2018A&A...616A...1G} {616, A1}

\bibitem[\protect\citeauthoryear{{Gaia Collaboration} et~al.,}{{Gaia
  Collaboration} et~al.}{2018b}]{gaia-hr2018}
{Gaia Collaboration} et~al., 2018b, \mn@doi [\aap]
  {10.1051/0004-6361/201832843}, \href
  {https://ui.adsabs.harvard.edu/abs/2018A&A...616A..10G} {616, A10}

\bibitem[\protect\citeauthoryear{{Gaia Collaboration}, {Brown}, {Vallenari},
  {Prusti}, {de Bruijne}, {Babusiaux}  \& {Biermann}}{{Gaia Collaboration}
  et~al.}{2020}]{Gaia2020}
{Gaia Collaboration} {Brown} A.~G.~A.,  {Vallenari} A.,  {Prusti} T.,  {de
  Bruijne} J.~H.~J.,  {Babusiaux} C.,   {Biermann} M.,  2020, arXiv e-prints,
  \href {https://ui.adsabs.harvard.edu/abs/2020arXiv201201533G} {p.
  arXiv:2012.01533}

\bibitem[\protect\citeauthoryear{{Gentile Fusillo} et~al.,}{{Gentile Fusillo}
  et~al.}{2019}]{gentile-fusillo2019}
{Gentile Fusillo} N.~P.,  et~al., 2019, \mn@doi [\mnras]
  {10.1093/mnras/sty3016}, \href
  {https://ui.adsabs.harvard.edu/abs/2019MNRAS.482.4570G} {482, 4570}

\bibitem[\protect\citeauthoryear{{Hayden} et~al.,}{{Hayden}
  et~al.}{2020}]{Hayden2020}
{Hayden} M.~R.,  et~al., 2020, arXiv e-prints, \href
  {https://ui.adsabs.harvard.edu/abs/2020arXiv201113745H} {p. arXiv:2011.13745}

\bibitem[\protect\citeauthoryear{{Haywood}}{{Haywood}}{2008}]{Haywood2008}
{Haywood} M.,  2008, \mn@doi [\mnras] {10.1111/j.1365-2966.2008.13395.x}, \href
  {https://ui.adsabs.harvard.edu/abs/2008MNRAS.388.1175H} {388, 1175}

\bibitem[\protect\citeauthoryear{{Haywood}, {Di Matteo}, {Lehnert}, {Katz}  \&
  {G{\'o}mez}}{{Haywood} et~al.}{2013}]{Haywood13}
{Haywood} M.,  {Di Matteo} P.,  {Lehnert} M.~D.,  {Katz} D.,   {G{\'o}mez} A.,
  2013, \mn@doi [\aap] {10.1051/0004-6361/201321397}, \href
  {https://ui.adsabs.harvard.edu/abs/2013A&A...560A.109H} {560, A109}

\bibitem[\protect\citeauthoryear{{Holtzman} et~al.,}{{Holtzman}
  et~al.}{2015}]{Holtzman2015}
{Holtzman} J.~A.,  et~al., 2015, \mn@doi [\aj] {10.1088/0004-6256/150/5/148},
  \href {http://adsabs.harvard.edu/abs/2015AJ....150..148H} {150, 148}

\bibitem[\protect\citeauthoryear{{Howes}, {Lindegren}, {Feltzing}, {Church}  \&
  {Bensby}}{{Howes} et~al.}{2019}]{Howes19}
{Howes} L.~M.,  {Lindegren} L.,  {Feltzing} S.,  {Church} R.~P.,   {Bensby} T.,
   2019, \mn@doi [\aap] {10.1051/0004-6361/201833280}, \href
  {https://ui.adsabs.harvard.edu/abs/2019A&A...622A..27H} {622, A27}

\bibitem[\protect\citeauthoryear{{Jim{\'e}nez-Esteban}, {Solano}  \&
  {Rodrigo}}{{Jim{\'e}nez-Esteban} et~al.}{2019}]{jimenez-esteban2019}
{Jim{\'e}nez-Esteban} F.~M.,  {Solano} E.,   {Rodrigo} C.,  2019, \mn@doi [\aj]
  {10.3847/1538-3881/aafacc}, \href
  {https://ui.adsabs.harvard.edu/abs/2019AJ....157...78J} {157, 78}

\bibitem[\protect\citeauthoryear{{J{\o}rgensen} \& {Lindegren}}{{J{\o}rgensen}
  \& {Lindegren}}{2005}]{Jorgensen05}
{J{\o}rgensen} B.~R.,  {Lindegren} L.,  2005, \mn@doi [\aap]
  {10.1051/0004-6361:20042185}, \href
  {https://ui.adsabs.harvard.edu/abs/2005A&A...436..127J} {436, 127}

\bibitem[\protect\citeauthoryear{{Koester}}{{Koester}}{2010}]{Koester2010}
{Koester} D.,  2010, \memsai, \href
  {https://ui.adsabs.harvard.edu/abs/2010MmSAI..81..921K} {81, 921}

\bibitem[\protect\citeauthoryear{{K{\"o}ppen} \& {Hensler}}{{K{\"o}ppen} \&
  {Hensler}}{2005}]{Koppen+Hensler2005}
{K{\"o}ppen} J.,  {Hensler} G.,  2005, \mn@doi [\aap]
  {10.1051/0004-6361:20042266}, \href
  {https://ui.adsabs.harvard.edu/abs/2005A&A...434..531K} {434, 531}

\bibitem[\protect\citeauthoryear{{Kurucz}}{{Kurucz}}{1993}]{kurucz93}
{Kurucz} R.,  1993, ATLAS9 Stellar Atmosphere Programs and 2 km/s grid. Kurucz
  CD-ROM No. 13. Cambridge, \href
  {https://ui.adsabs.harvard.edu/abs/1993KurCD..13.....K} {13}

\bibitem[\protect\citeauthoryear{{Lallement}, {Vergely}, {Valette},
  {Puspitarini}, {Eyer}  \& {Casagrande}}{{Lallement}
  et~al.}{2014}]{Lallement2014}
{Lallement} R.,  {Vergely} J.~L.,  {Valette} B.,  {Puspitarini} L.,  {Eyer} L.,
    {Casagrande} L.,  2014, \mn@doi [\aap] {10.1051/0004-6361/201322032}, \href
  {https://ui.adsabs.harvard.edu/abs/2014A&A...561A..91L} {561, A91}

\bibitem[\protect\citeauthoryear{{Lam}, {Hambly}, {Lodieu}, {Blouin}, {Harvey},
  {Smith}, {G{\'a}lvez-Ortiz}  \& {Zhang}}{{Lam} et~al.}{2020}]{Lam2020}
{Lam} M.~C.,  {Hambly} N.~C.,  {Lodieu} N.,  {Blouin} S.,  {Harvey} E.~J.,
  {Smith} R.~J.,  {G{\'a}lvez-Ortiz} M.~C.,   {Zhang} Z.~H.,  2020, \mn@doi
  [\mnras] {10.1093/mnras/staa584}, \href
  {https://ui.adsabs.harvard.edu/abs/2020MNRAS.493.6001L} {493, 6001}

\bibitem[\protect\citeauthoryear{Lindegren}{Lindegren}{2018}]{lindegren2018-b}
Lindegren L.,  2018, Technical Report GAIA-C3-TN-LU-LL-124-01, Re-normalising
  the astrometric chi-square in {\em Gaia} DR2, \url
  {http://www.rssd.esa.int/doc_fetch.php?id=3757412}.
Lund Observatory, Lund, Sweden, \url
  {http://www.rssd.esa.int/doc_fetch.php?id=3757412}

\bibitem[\protect\citeauthoryear{{Lindgren} \& {Heiter}}{{Lindgren} \&
  {Heiter}}{2017}]{Lindgren2017}
{Lindgren} S.,  {Heiter} U.,  2017, \mn@doi [\aap]
  {10.1051/0004-6361/201730715}, \href
  {https://ui.adsabs.harvard.edu/abs/2017A&A...604A..97L} {604, A97}

\bibitem[\protect\citeauthoryear{{Maldonado} et~al.,}{{Maldonado}
  et~al.}{2015}]{Maldonado2015}
{Maldonado} J.,  et~al., 2015, \mn@doi [\aap] {10.1051/0004-6361/201525797},
  \href {https://ui.adsabs.harvard.edu/abs/2015A&A...577A.132M} {577, A132}

\bibitem[\protect\citeauthoryear{{Marsh}}{{Marsh}}{1989}]{Marsh1989}
{Marsh} T.~R.,  1989, \mn@doi [\pasp] {10.1086/132570}, \href
  {https://ui.adsabs.harvard.edu/abs/1989PASP..101.1032M} {101, 1032}

\bibitem[\protect\citeauthoryear{{McCleery} et~al.,}{{McCleery}
  et~al.}{2020}]{McCleery2020}
{McCleery} J.,  et~al., 2020, \mn@doi [\mnras] {10.1093/mnras/staa2030}, \href
  {https://ui.adsabs.harvard.edu/abs/2020MNRAS.499.1890M} {499, 1890}

\bibitem[\protect\citeauthoryear{{Miglio} et~al.,}{{Miglio}
  et~al.}{2021}]{Miglio2021}
{Miglio} A.,  et~al., 2021, \mn@doi [\aap] {10.1051/0004-6361/202038307}, \href
  {https://ui.adsabs.harvard.edu/abs/2021A&A...645A..85M} {645, A85}

\bibitem[\protect\citeauthoryear{{Miller Bertolami}}{{Miller
  Bertolami}}{2016}]{Miller2016}
{Miller Bertolami} M.~M.,  2016, \mn@doi [\aap] {10.1051/0004-6361/201526577},
  \href {https://ui.adsabs.harvard.edu/abs/2016A&A...588A..25M} {588, A25}

\bibitem[\protect\citeauthoryear{{Minchev}, {Famaey}, {Combes}, {Di Matteo},
  {Mouhcine}  \& {Wozniak}}{{Minchev} et~al.}{2011}]{Minchev2011}
{Minchev} I.,  {Famaey} B.,  {Combes} F.,  {Di Matteo} P.,  {Mouhcine} M.,
  {Wozniak} H.,  2011, \mn@doi [\aap] {10.1051/0004-6361/201015139}, \href
  {https://ui.adsabs.harvard.edu/abs/2011A&A...527A.147M} {527, A147}

\bibitem[\protect\citeauthoryear{{Minchev}, {Chiappini}  \& {Martig}}{{Minchev}
  et~al.}{2013}]{Minchev2013}
{Minchev} I.,  {Chiappini} C.,   {Martig} M.,  2013, \mn@doi [\aap]
  {10.1051/0004-6361/201220189}, \href
  {https://ui.adsabs.harvard.edu/abs/2013A&A...558A...9M} {558, A9}

\bibitem[\protect\citeauthoryear{{Nebot G{\'o}mez-Mor{\'a}n} et~al.,}{{Nebot
  G{\'o}mez-Mor{\'a}n} et~al.}{2011}]{Nebot2011}
{Nebot G{\'o}mez-Mor{\'a}n} A.,  et~al., 2011, \mn@doi [\aap]
  {10.1051/0004-6361/201117514}, \href
  {https://ui.adsabs.harvard.edu/abs/2011A&A...536A..43N} {536, A43}

\bibitem[\protect\citeauthoryear{{Newton}, {Charbonneau}, {Irwin},
  {Berta-Thompson}, {Rojas-Ayala}, {Covey}  \& {Lloyd}}{{Newton}
  et~al.}{2014}]{Newton2014}
{Newton} E.~R.,  {Charbonneau} D.,  {Irwin} J.,  {Berta-Thompson} Z.~K.,
  {Rojas-Ayala} B.,  {Covey} K.,   {Lloyd} J.~P.,  2014, \mn@doi [\aj]
  {10.1088/0004-6256/147/1/20}, \href
  {https://ui.adsabs.harvard.edu/abs/2014AJ....147...20N} {147, 20}

\bibitem[\protect\citeauthoryear{{Nissen}, {Christensen-Dalsgaard},
  {Mosumgaard}, {Silva Aguirre}, {Spitoni}  \& {Verma}}{{Nissen}
  et~al.}{2020}]{Nissen2020}
{Nissen} P.~E.,  {Christensen-Dalsgaard} J.,  {Mosumgaard} J.~R.,  {Silva
  Aguirre} V.,  {Spitoni} E.,   {Verma} K.,  2020, \mn@doi [\aap]
  {10.1051/0004-6361/202038300}, \href
  {https://ui.adsabs.harvard.edu/abs/2020A&A...640A..81N} {640, A81}

\bibitem[\protect\citeauthoryear{{Nomoto}, {Kobayashi}  \& {Tominaga}}{{Nomoto}
  et~al.}{2013}]{Nomoto13}
{Nomoto} K.,  {Kobayashi} C.,   {Tominaga} N.,  2013, \mn@doi [\araa]
  {10.1146/annurev-astro-082812-140956}, \href
  {https://ui.adsabs.harvard.edu/abs/2013ARA&A..51..457N} {51, 457}

\bibitem[\protect\citeauthoryear{{Nordstr{\"o}m} et~al.,}{{Nordstr{\"o}m}
  et~al.}{2004}]{Nordstrom2004}
{Nordstr{\"o}m} B.,  et~al., 2004, \mn@doi [\aap] {10.1051/0004-6361:20035959},
  \href {https://ui.adsabs.harvard.edu/abs/2004A&A...418..989N} {418, 989}

\bibitem[\protect\citeauthoryear{{Pancino}, {Carrera}, {Rossetti}  \&
  {Gallart}}{{Pancino} et~al.}{2010}]{Pancino2010}
{Pancino} E.,  {Carrera} R.,  {Rossetti} E.,   {Gallart} C.,  2010, \mn@doi
  [\aap] {10.1051/0004-6361/200912965}, \href
  {https://ui.adsabs.harvard.edu/abs/2010A&A...511A..56P} {511, A56}

\bibitem[\protect\citeauthoryear{{Parsons} et~al.,}{{Parsons}
  et~al.}{2012}]{Parsons2012}
{Parsons} S.~G.,  et~al., 2012, \mn@doi [\mnras]
  {10.1111/j.1365-2966.2012.21773.x}, \href
  {https://ui.adsabs.harvard.edu/abs/2012MNRAS.426.1950P} {426, 1950}

\bibitem[\protect\citeauthoryear{{Pickles}}{{Pickles}}{1998}]{pickles98}
{Pickles} A.~J.,  1998, \mn@doi [\pasp] {10.1086/316197}, \href
  {https://ui.adsabs.harvard.edu/abs/1998PASP..110..863P} {110, 863}

\bibitem[\protect\citeauthoryear{{Pilyugin} \& {Edmunds}}{{Pilyugin} \&
  {Edmunds}}{1996}]{Pilyugin+Edmunds1996}
{Pilyugin} L.~S.,  {Edmunds} M.~G.,  1996, \aap, \href
  {https://ui.adsabs.harvard.edu/abs/1996A&A...313..792P} {313, 792}

\bibitem[\protect\citeauthoryear{{Pinsonneault} et~al.,}{{Pinsonneault}
  et~al.}{2018}]{Pinsonneault18}
{Pinsonneault} M.~H.,  et~al., 2018, \mn@doi [\apjs]
  {10.3847/1538-4365/aaebfd}, \href
  {https://ui.adsabs.harvard.edu/abs/2018ApJS..239...32P} {239, 32}

\bibitem[\protect\citeauthoryear{{Qiu}, {Tian}, {Wang}, {Nie}, {von Hippe},
  {Liu}, {Fouesneau}  \& {Rix}}{{Qiu} et~al.}{2020}]{Qiu2020}
{Qiu} D.,  {Tian} H.-J.,  {Wang} X.-D.,  {Nie} J.-L.,  {von Hippe} T.,  {Liu}
  G.-C.,  {Fouesneau} M.,   {Rix} H.-W.,  2020, arXiv e-prints, \href
  {https://ui.adsabs.harvard.edu/abs/2020arXiv201204890Q} {p. arXiv:2012.04890}

\bibitem[\protect\citeauthoryear{{Raskin} et~al.,}{{Raskin}
  et~al.}{2011}]{Raskin2011}
{Raskin} G.,  et~al., 2011, \mn@doi [\aap] {10.1051/0004-6361/201015435}, \href
  {https://ui.adsabs.harvard.edu/abs/2011A&A...526A..69R} {526, A69}

\bibitem[\protect\citeauthoryear{{Rebassa-Mansergas}, {G{\"a}nsicke},
  {Rodr{\'\i}guez-Gil}, {Schreiber}  \& {Koester}}{{Rebassa-Mansergas}
  et~al.}{2007}]{Rebassa2007}
{Rebassa-Mansergas} A.,  {G{\"a}nsicke} B.~T.,  {Rodr{\'\i}guez-Gil} P.,
  {Schreiber} M.~R.,   {Koester} D.,  2007, \mn@doi [\mnras]
  {10.1111/j.1365-2966.2007.12288.x}, \href
  {https://ui.adsabs.harvard.edu/abs/2007MNRAS.382.1377R} {382, 1377}

\bibitem[\protect\citeauthoryear{{Rebassa-Mansergas}, {G{\"a}nsicke},
  {Schreiber}, {Koester}  \& {Rodr{\'\i}guez-Gil}}{{Rebassa-Mansergas}
  et~al.}{2010}]{Rebassa2010}
{Rebassa-Mansergas} A.,  {G{\"a}nsicke} B.~T.,  {Schreiber} M.~R.,  {Koester}
  D.,   {Rodr{\'\i}guez-Gil} P.,  2010, \mn@doi [\mnras]
  {10.1111/j.1365-2966.2009.15915.x}, \href
  {https://ui.adsabs.harvard.edu/abs/2010MNRAS.402..620R} {402, 620}

\bibitem[\protect\citeauthoryear{{Rebassa-Mansergas}, {Nebot
  G{\'o}mez-Mor{\'a}n}, {Schreiber}, {Girven}  \&
  {G{\"a}nsicke}}{{Rebassa-Mansergas} et~al.}{2011}]{Rebassa2011}
{Rebassa-Mansergas} A.,  {Nebot G{\'o}mez-Mor{\'a}n} A.,  {Schreiber} M.~R.,
  {Girven} J.,   {G{\"a}nsicke} B.~T.,  2011, \mn@doi [\mnras]
  {10.1111/j.1365-2966.2011.18200.x}, \href
  {https://ui.adsabs.harvard.edu/abs/2011MNRAS.413.1121R} {413, 1121}

\bibitem[\protect\citeauthoryear{{Rebassa-Mansergas}, {Ren}, {Parsons},
  {G{\"a}nsicke}, {Schreiber}, {Garc{\'\i}a-Berro}, {Liu}  \&
  {Koester}}{{Rebassa-Mansergas} et~al.}{2016a}]{Rebassa2016b}
{Rebassa-Mansergas} A.,  {Ren} J.~J.,  {Parsons} S.~G.,  {G{\"a}nsicke} B.~T.,
  {Schreiber} M.~R.,  {Garc{\'\i}a-Berro} E.,  {Liu} X.~W.,   {Koester} D.,
  2016a, \mn@doi [\mnras] {10.1093/mnras/stw554}, \href
  {https://ui.adsabs.harvard.edu/abs/2016MNRAS.458.3808R} {458, 3808}

\bibitem[\protect\citeauthoryear{{Rebassa-Mansergas}
  et~al.,}{{Rebassa-Mansergas} et~al.}{2016b}]{Rebassa16AMR}
{Rebassa-Mansergas} A.,  et~al., 2016b, \mn@doi [\mnras]
  {10.1093/mnras/stw2021}, \href
  {https://ui.adsabs.harvard.edu/abs/2016MNRAS.463.1137R} {463, 1137}

\bibitem[\protect\citeauthoryear{{Renedo}, {Althaus}, {Miller Bertolami},
  {Romero}, {C{\'o}rsico}, {Rohrmann}  \& {Garc{\'\i}a-Berro}}{{Renedo}
  et~al.}{2010}]{Renedo2010}
{Renedo} I.,  {Althaus} L.~G.,  {Miller Bertolami} M.~M.,  {Romero} A.~D.,
  {C{\'o}rsico} A.~H.,  {Rohrmann} R.~D.,   {Garc{\'\i}a-Berro} E.,  2010,
  \mn@doi [\apj] {10.1088/0004-637X/717/1/183}, \href
  {https://ui.adsabs.harvard.edu/abs/2010ApJ...717..183R} {717, 183}

\bibitem[\protect\citeauthoryear{{Riello} et~al.,}{{Riello}
  et~al.}{2020}]{Riello2020}
{Riello} M.,  et~al., 2020, arXiv e-prints, \href
  {https://ui.adsabs.harvard.edu/abs/2020arXiv201201916R} {p. arXiv:2012.01916}

\bibitem[\protect\citeauthoryear{{Rocha-Pinto}, {Maciel}, {Scalo}  \&
  {Flynn}}{{Rocha-Pinto} et~al.}{2000}]{RochaPinto2000}
{Rocha-Pinto} H.~J.,  {Maciel} W.~J.,  {Scalo} J.,   {Flynn} C.,  2000, \aap,
  \href {https://ui.adsabs.harvard.edu/abs/2000A&A...358..850R} {358, 850}

\bibitem[\protect\citeauthoryear{{Ro{\v{s}}kar}, {Debattista}, {Quinn},
  {Stinson}  \& {Wadsley}}{{Ro{\v{s}}kar} et~al.}{2008}]{Roskar2008}
{Ro{\v{s}}kar} R.,  {Debattista} V.~P.,  {Quinn} T.~R.,  {Stinson} G.~S.,
  {Wadsley} J.,  2008, \mn@doi [\apjl] {10.1086/592231}, \href
  {https://ui.adsabs.harvard.edu/abs/2008ApJ...684L..79R} {684, L79}

\bibitem[\protect\citeauthoryear{{Sellwood} \& {Binney}}{{Sellwood} \&
  {Binney}}{2002}]{SellwoodBinney2002}
{Sellwood} J.~A.,  {Binney} J.~J.,  2002, \mn@doi [\mnras]
  {10.1046/j.1365-8711.2002.05806.x}, \href
  {https://ui.adsabs.harvard.edu/abs/2002MNRAS.336..785S} {336, 785}

\bibitem[\protect\citeauthoryear{{Silva Aguirre} et~al.,}{{Silva Aguirre}
  et~al.}{2018}]{SilvaAguirre2018}
{Silva Aguirre} V.,  et~al., 2018, \mn@doi [\mnras] {10.1093/mnras/sty150},
  \href {https://ui.adsabs.harvard.edu/abs/2018MNRAS.475.5487S} {475, 5487}

\bibitem[\protect\citeauthoryear{{Soubiran}, {Bienaym{\'e}}, {Mishenina}  \&
  {Kovtyukh}}{{Soubiran} et~al.}{2008}]{Soubiran08}
{Soubiran} C.,  {Bienaym{\'e}} O.,  {Mishenina} T.~V.,   {Kovtyukh} V.~V.,
  2008, \mn@doi [\aap] {10.1051/0004-6361:20078788}, \href
  {https://ui.adsabs.harvard.edu/abs/2008A&A...480...91S} {480, 91}

\bibitem[\protect\citeauthoryear{{Takeda}, {Ohkubo}, {Sato}, {Kambe}  \&
  {Sadakane}}{{Takeda} et~al.}{2005}]{takeda05}
{Takeda} Y.,  {Ohkubo} M.,  {Sato} B.,  {Kambe} E.,   {Sadakane} K.,  2005,
  \mn@doi [\pasj] {10.1093/pasj/57.1.27}, \href
  {https://ui.adsabs.harvard.edu/abs/2005PASJ...57...27T} {57, 27}

\bibitem[\protect\citeauthoryear{{Teixeira}, {Sousa}, {Tsantaki}, {Monteiro},
  {Santos}  \& {Israelian}}{{Teixeira} et~al.}{2016}]{Teixeira2016}
{Teixeira} G.~D.~C.,  {Sousa} S.~G.,  {Tsantaki} M.,  {Monteiro}
  M.~J.~P.~F.~G.,  {Santos} N.~C.,   {Israelian} G.,  2016, \mn@doi [\aap]
  {10.1051/0004-6361/201525783}, \href
  {https://ui.adsabs.harvard.edu/abs/2016A&A...595A..15T} {595, A15}

\bibitem[\protect\citeauthoryear{{Tody}}{{Tody}}{1986}]{Tody1986}
{Tody} D.,  1986, in {Crawford} D.~L.,  ed.,  Society of Photo-Optical
  Instrumentation Engineers (SPIE) Conference Series Vol. 627, Instrumentation
  in astronomy VI. p.~733, \mn@doi{10.1117/12.968154}

\bibitem[\protect\citeauthoryear{{Torres}, {Rebassa-Mansergas}, {Camisassa}  \&
  {Raddi}}{{Torres} et~al.}{2021}]{Torres2021}
{Torres} S.,  {Rebassa-Mansergas} A.,  {Camisassa} M.~E.,   {Raddi} R.,  2021,
  \mn@doi [\mnras] {10.1093/mnras/stab079}, \href
  {https://ui.adsabs.harvard.edu/abs/2021MNRAS.502.1753T} {502, 1753}

\bibitem[\protect\citeauthoryear{{Tremblay}, {Ludwig}, {Steffen}  \&
  {Freytag}}{{Tremblay} et~al.}{2013}]{Tremblay2013}
{Tremblay} P.~E.,  {Ludwig} H.~G.,  {Steffen} M.,   {Freytag} B.,  2013,
  \mn@doi [\aap] {10.1051/0004-6361/201322318}, \href
  {https://ui.adsabs.harvard.edu/abs/2013A&A...559A.104T} {559, A104}

\bibitem[\protect\citeauthoryear{{Tremblay} et~al.,}{{Tremblay}
  et~al.}{2020}]{Tremblay2020}
{Tremblay} P.~E.,  et~al., 2020, \mn@doi [\mnras] {10.1093/mnras/staa1892},
  \href {https://ui.adsabs.harvard.edu/abs/2020MNRAS.497..130T} {497, 130}

\bibitem[\protect\citeauthoryear{{Twarog}}{{Twarog}}{1980}]{Twarog80}
{Twarog} B.~A.,  1980, \mn@doi [\apj] {10.1086/158460}, \href
  {https://ui.adsabs.harvard.edu/abs/1980ApJ...242..242T} {242, 242}

\bibitem[\protect\citeauthoryear{{Vernet} et~al.,}{{Vernet}
  et~al.}{2011}]{Vernet2011}
{Vernet} J.,  et~al., 2011, \mn@doi [\aap] {10.1051/0004-6361/201117752}, \href
  {https://ui.adsabs.harvard.edu/abs/2011A&A...536A.105V} {536, A105}

\bibitem[\protect\citeauthoryear{{Willems} \& {Kolb}}{{Willems} \&
  {Kolb}}{2004}]{Willems04}
{Willems} B.,  {Kolb} U.,  2004, \mn@doi [\aap] {10.1051/0004-6361:20040085},
  \href {https://ui.adsabs.harvard.edu/abs/2004A&A...419.1057W} {419, 1057}

\bibitem[\protect\citeauthoryear{{Wojno} et~al.,}{{Wojno}
  et~al.}{2018}]{Wojno2018}
{Wojno} J.,  et~al., 2018, \mn@doi [\mnras] {10.1093/mnras/sty1016}, \href
  {https://ui.adsabs.harvard.edu/abs/2018MNRAS.477.5612W} {477, 5612}

\bibitem[\protect\citeauthoryear{{Wu}, {Xiang}, {Chen}, {Zhao}, {Bi}, {Li},
  {Li}  \& {Huang}}{{Wu} et~al.}{2021}]{Wu2021}
{Wu} Y.,  {Xiang} M.,  {Chen} Y.,  {Zhao} G.,  {Bi} S.,  {Li} C.,  {Li} Y.,
  {Huang} Y.,  2021, \mn@doi [\mnras] {10.1093/mnras/staa3949}, \href
  {https://ui.adsabs.harvard.edu/abs/2021MNRAS.501.4917W} {501, 4917}

\bibitem[\protect\citeauthoryear{{York} et~al.,}{{York}
  et~al.}{2000}]{York2000}
{York} D.~G.,  et~al., 2000, \mn@doi [\aj] {10.1086/301513}, \href
  {https://ui.adsabs.harvard.edu/abs/2000AJ....120.1579Y} {120, 1579}

\bibitem[\protect\citeauthoryear{{Yuan}, {Liu}, {Xiang}, {Huang}, {Chen}, {Wu},
  {Hou}  \& {Zhang}}{{Yuan} et~al.}{2015}]{Yuan15}
{Yuan} H.,  {Liu} X.,  {Xiang} M.,  {Huang} Y.,  {Chen} B.,  {Wu} Y.,  {Hou}
  Y.,   {Zhang} Y.,  2015, \mn@doi [\apj] {10.1088/0004-637X/799/2/135}, \href
  {https://ui.adsabs.harvard.edu/abs/2015ApJ...799..135Y} {799, 135}

\makeatother
\end{thebibliography}



\bsp	
\label{lastpage}
\end{document}